\documentclass[fleqn,usenatbib]{mnras}

\usepackage{newtxtext,newtxmath}
\usepackage[T1]{fontenc}
\DeclareRobustCommand{\VAN}[3]{#2}
\let\VANthebibliography\thebibliography
\def\thebibliography{\DeclareRobustCommand{\VAN}[3]{##3}\VANthebibliography}

\usepackage{graphicx}	
\usepackage{amsmath}	
\usepackage{caption}
\usepackage{subcaption}
\usepackage{placeins}
\usepackage{hyperref}
\usepackage{rotating}
\usepackage{CJK}

\newcommand{\teff}{$T_{\rm eff}$} 
\newcommand{\logg}{$\log g$}

\newcommand\msun{M$_{\odot}$}

\title[SDST II. Design, observations and data]{Survey for Distant Solar Twins (SDST) -- II. Design, observations and data}
\author[F.\ Liu et al.]
{Fan Liu (刘凡),$^{1,2,3}$\thanks{E-mail: fan.liu@monash.edu}
Michael T.\ Murphy,$^{1}$
Christian Lehmann,$^{1}$
Chris Flynn,$^{1}$
Daniel Smith,$^{1,4}$
Janez Kos,$^{5}$ \newauthor
Daniel A.\ Berke$^{1,6}$ and
Sarah L.\ Martell$^{3,7}$ \\
\\
$^{1}$Centre for Astrophysics and Supercomputing, Swinburne University of Technology, Melbourne, VIC 3122, Australia \\
$^{2}$School of Physics and Astronomy, Monash University, Melbourne, VIC 3800, Australia \\
$^{3}$ARC Centre for All Sky Astrophysics in 3D (ASTRO-3D), Canberra, ACT 0200, Australia \\
$^{4}$Optical Sciences Centre, Swinburne University of Technology, Melbourne, VIC 3122, Australia \\
$^{5}$Faculty of Mathematics and Physics, University of Ljubljana, Jadranska 19, 1000 Ljubljana, Slovenia \\
$^{6}$User Support Division, Gemini Observatory, 670 N A'ohoku Pl, Hilo, HI 96720-2700, USA \\
$^{7}$School of Physics, University of New South Wales, Sydney, NSW 2052, Australia
}

\date{Accepted 2022 October 17. Received 2022 October 12; in original form 2022 September 27}

\pubyear{2022}

\begin{document}
\begin{CJK*}{UTF8}{gbsn}
\label{firstpage}
\pagerange{\pageref{firstpage}--\pageref{lastpage}}
\maketitle

\begin{abstract}
Studies of solar twins have key impacts on the astronomical community, but only $\sim$ 100--200 nearby solar twins ($<$ 1\,kpc) have been reliably identified over the last few decades. The aim of our survey (SDST) is to identify $\sim$\,150--200 distant solar twins and analogues (up to $\lesssim$ 4\,kpc) closer to the Galactic Centre. We took advantage of the precise Gaia and Skymapper surveys to select Sun-like candidates in a 2-degree field, which were observed with the HERMES spectrograph on the Anglo-Australian Telescope. We successfully built up the required signal-to-noise ratio (25-per-pixel in the HERMES red band) for most targets as faint as Gaia G of 17.4\,mag. The stellar photometric/astrometric parameters (e.g., \teff, \logg, mass) of our candidates are derived in this paper, while the spectroscopic parameters will be presented in the third paper in this SDST series. The selection success rate – the fraction of targets which belong to solar twins or analogues – was estimated from simulated survey data and the Besan\c{c}on stellar population model, and compared with the actual success rate of the survey. We find that expected and actual success rates agree well, indicating that the numbers of solar twins and analogues we discover in SDST are consistent with expectations, affirming the survey approach. These distant solar analogues are prime targets for testing for any variation in the strength of electromagnetism in regions of higher dark matter density, and can make additional contributions to our understanding of, e.g., Galactic chemical evolution in the inner Milky Way.
\end{abstract}

\begin{keywords}
methods: observational -- stars: fundamental parameters -- stars: general -- stars: solar-type
\end{keywords}




\section{Introduction}

Detection of solar twins, i.e. stars with similar physical parameters to the Sun, is one of the fundamental tasks in astronomical community over the last few decades, because these stars have been serving as the standard reference sample for various applications. The early attempt to find solar twins was conducted by \citet{har78}, and the first solar twin was identified by \citet{pd97}. Later studies have analysed and identified $\sim$\,100 -- 200 nearby solar twins based on their spectroscopic and/or photometric parameters, using different definitions of what constitutes a true "twin" (e.g., \citealp{kin05,mel06,mr07,mel14a,bau10,dat12,dat14,dat15,por14,ram14,gal16,gal21}). Most of them are within 500\,pc from the Sun, and the most distant solar twins were found in the M67 open cluster \citep{pas08,one11,liu16} with a distance of 890\,pc \citep{kha13}. 

Studies of these local solar twins have made significant contributions to different fields of astrophysics. For example, they enable determination of the zero-point for fundamental photometric calibrations \citep{hol06,cas10,cas21,ram12}; better understanding of the chemical evolution of the Galactic disc \citep{spi16,bed18,bot20}; the nucleosynthetic history and chemical clocks \citep{mel14b,nis15,nis20,spi18}, as well as the important lithium-age correlation \citep{car16,car19}; the peculiar chemical composition of the Sun \citep{mel09,ram09,ram10} and its implication for the Sun's birth environment \citep{one11,liu16,gus18}; and building a connection between accurate stellar and planetary properties (e.g., \citealp{wang19,liu20,adi21}). In addition, solar twins can serve as new probes to test for physics beyond the Standard Model of particle physics, as introduced in \citet{mur22a} and \citet{ber22a} -- we briefly summarise this new application below. 

\begin{figure}
\centering
\includegraphics[width=\columnwidth]{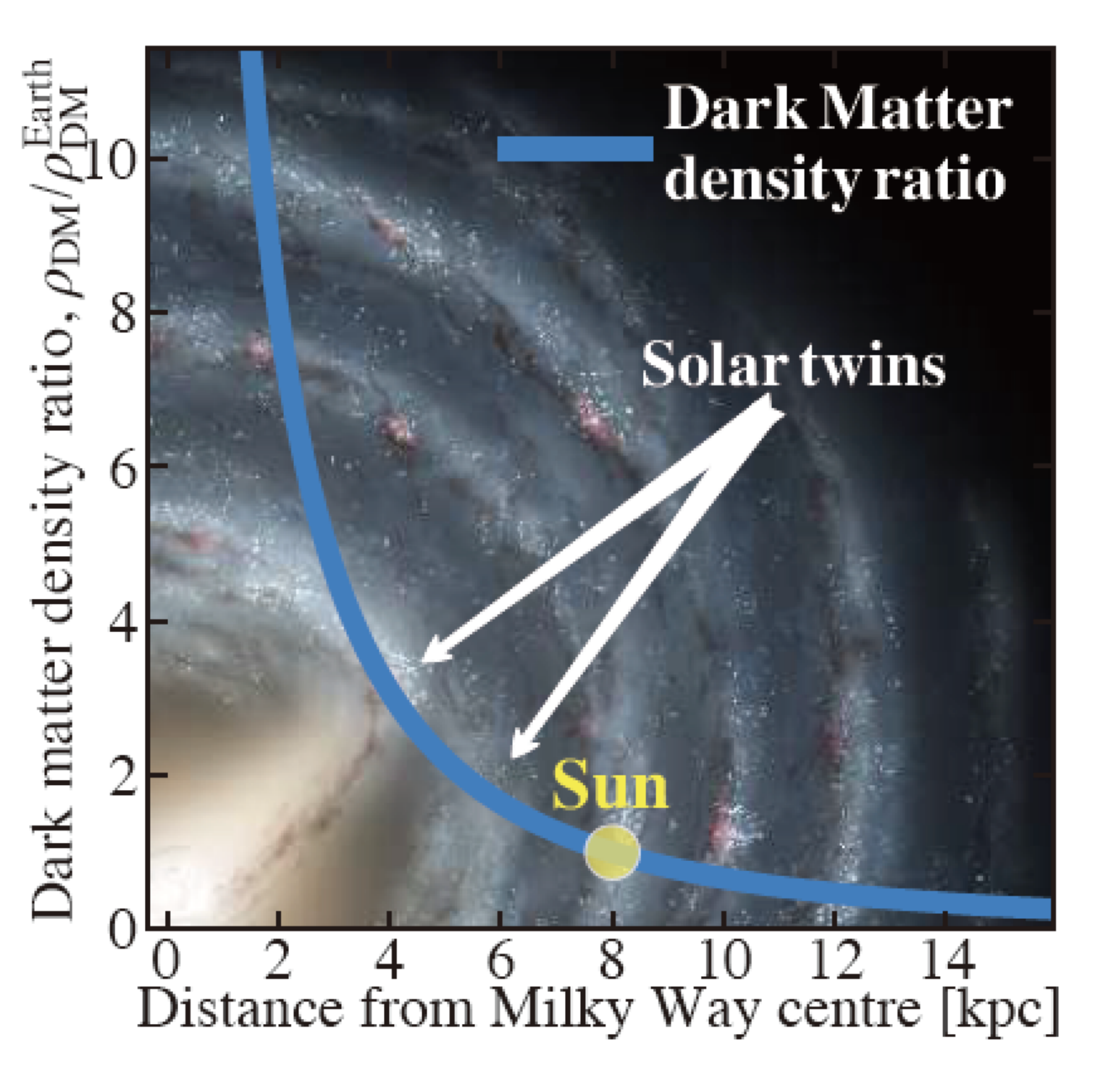}
\caption{Illustration of the survey goal. Dark Matter density varies strongly across our galaxy so we aim to discover $\sim$\,150--200 distant solar twins and analogues ($\lesssim$ 4\,kpc) closer to the Galactic Centre, as new probes of variation in $\alpha$ against varying Dark Matter density.}
\label{fig:motivation}
\end{figure}

The predictions of the Standard Model have been tested with extreme, 1\,part-per-billion (ppb) precision in Earth-bound laboratories \citep{han08,aoy15}. However, the foundations of the Standard Model are incomplete: it contains numerous "fundamental constants" such as $\alpha$, the indicator of electromagnetism's strength, but cannot explain their values, origin, constancy or otherwise. Therefore, precise searches for variations in the fine-structure constant $\alpha$ probe new, "beyond-Standard" physics. Short-term variability has been tested with atomic clocks \citep{ros08,lan21}, and cosmological time/space variations have been limited to 1\,part-per-million (ppm) by quasar spectroscopy \citep{mur17,mur22b}. However, $\alpha$'s constancy has never been directly and precisely mapped across our own Galaxy, where Dark Matter density varies strongly (see Fig.~\ref{fig:motivation}). The Standard Model has no explanation for Dark Matter, and some theories predict it may be linked with variations in $\alpha$ (examples \citealp{oli02,sta15,eic18,dav19}). Therefore, mapping $\alpha$ across the Milky Way would be a completely new test of fundamental physics. 

Until now, this was impossible due to the lack of precise Galactic probes of $\alpha$ with well-controlled systematic errors. Atomic line wavelengths have different sensitivities to $\alpha$, so comparing multiple lines to their laboratory wavelengths directly probes $\alpha$ variation. While this basic principle was recently applied to stellar spectra \citep{hee20}, line shifts and asymmetries from astrophysical processes in the stellar photospheres cause strong systematics: they vary with line optical depth and stellar parameters, limiting $\alpha$ measurement accuracy to $\sim$ 10\,ppm (relative line shifts of $\sim$ 200\,m\,s$^{-1}$). This problem can be solved using solar twins and analogues as new probes by comparing separations between pairs of selected spectral lines in these intrinsically similar stars. 

This approach was illustrated and comprehensively tested using nearby Sun-like stars by \citet{ber22a,ber22b}. They carefully selected 229 pairs of relatively unblended transitions with similar optical depths, avoiding even weak telluric features, and compared their relative separations between these stars. The pairs are closely separated ($<$ 800\,km\,s$^{-1}$) to avoid instrumental systematic effects (e.g., wavelength calibration). This strictly differential approach strongly suppresses systematics, enabling 100-times-better accuracy for not only solar twins, but also Sun-like stars with a relatively broad range of stellar parameters around solar values. With this approach, \citet{mur22a} constrained relative variations in $\alpha$ between 17 local solar twins to $\lesssim$ 50\,ppb and set a local, empirical reference, against which more distant stars can be compared, with 12\,ppb precision. 

The definition of a solar twin varies between studies, and has evolved with time and available technology \citep{cay81,cay96,fri93,ram09,dat12,oli18,gal21}, but they all require atmospheric parameters very similar to the solar values (effective temperature \teff: 5772\,K, surface gravity \logg\footnote{$g$ in units of cm s$^{-2}$}: 4.44\,dex, and metallicity [Fe/H]: 0.0\,dex; \citealp{prs16}). Here we define "solar twins" as stars with (\teff, \logg, [Fe/H]) within $\pm$(100\,K, 0.2\,dex, 0.1\,dex) around the solar values. Stars with a larger range of physical parameters around solar values are usually defined as solar analogues or solar-type stars \citep{por14}. In this paper we define "solar analogues" as stars within $\pm$(300\,K, 0.4\,dex and 0.3\,dex) around the solar values. As illustrated in \citet{ber22a,ber22b}, the use of solar twins and analogues within such definition ensure that the majority of systematics are sufficiently suppressed or reliably corrected in a differential pair-separation analysis, enabling us to search for the potential $\alpha$ variations across our galaxy. 

To identify $\sim$\,150--200 solar twins and analogues much closer to the Galactic Centre, we have undertaken the Survey for Distant Solar Twins (SDST). The best $\approx$\,30 candidates from SDST will allow us to measure precisely the variations in $\alpha$ against the increasing Dark Matter density, as illustrated in Fig.~\ref{fig:motivation}. \citet[][hereafter SDST I]{leh22a} was the first paper in this series, which demonstrated a new method for measuring the stellar parameters of Sun-like stars from HERMES spectra. In this paper, SDST II, we describe our target selection and observations with HERMES. In SDST III (Lehmann et al., in prep.) we will present the stellar parameter measurements and a catalogue of the new solar twin and analogue discoveries. This paper is organised as follows: we present the survey design in Section 2, the observations and data reduction in Section 3, the survey verification in Section 4, and summarise the survey in Section 5.

\section{Survey design}

\subsection{Survey goal}

The goal of the survey is to discover and spectroscopically verify solar twins and analogues closer to the Galactic Centre, with distance up to 4\,kpc from us (see Fig.~\ref{fig:motivation}). The confirmed candidates can then be used to probe for variations in $\alpha$ with high precision (50--100\,ppb; see e.g., \citealp{mur22a}), which will enable us to test and map the $\alpha$ variations across regions with higher Dark Matter density nearer the Galactic Centre. These distant solar twins and analogues can be as faint as $\sim$ 17 -- 18\,mag, making it challenging to obtain their spectra with sufficient signal-to-noise ratio (SNR) for determination of their stellar parameters such as \teff, \logg, and [Fe/H]. 

Thanks to the ongoing large and deep photometric surveys of our galaxy and advancing instrumentation, we are now poised to achieve the survey goal. Several timely factors make it possible to discover distant solar twins and analogues, and probe $\alpha$ variations on a Galactic scale: \\
1. The data from the Gaia mission \citep{gaia16} and Skymapper Southern Sky Survey \citep{kel07} enable efficient selection of Sun-like candidates down to Gaia G $<$ 18 -- 19\,mag; \\
2. The High Efficiency and Resolution Multi-Element Spectrograph (HERMES; \citealp{des15,she15}) mounted on the Anglo-Australian Telescope (AAT) has the resolution and multiplex advantage for confirming candidates down to Gaia G $<$ 17.4\,mag with spectra of SNR $\geq$ 25 in the red band, based on the recently developed "\textsc{EPIC}" method for measuring the spectroscopic stellar parameters (SDST I); \\ 
3. The Echelle SPectrograph for Rocky Exoplanets and Stable Spectroscopic Observations (ESPRESSO; \citealp{pep21}) installed on the European Southern Observatory's Very Large Telescope (VLT) can provide high-resolution, high wavelength-accuracy spectra with sufficient SNR of the best HERMES-confirmed solar twin and analogue discoveries. It can provide the most precise determination of $\alpha$ across the Galactic regions with different Dark Matter density with the solar twins method established in \citet{mur22a} and \citet{ber22a,ber22b}. 

In order to achieve the survey goal, we targeted a 2-degree field with AAT/HERMES (see Section 2.2) and collected spectra of more than 850 Sun-like candidates selected based on their photometric and astrometric properties (see Section 2.3). We emphasize the requirement of the SNR of HERMES spectra to be $\geq$ 25 in the red band for our candidates, for which the \textsc{EPIC} method can efficiently and accurately measure their stellar atmospheric parameters for spectroscopic verification. Of the $\sim$\,150--200 twins/analogues we aim to discover, the slowest-rotating twins and analogues with no detectable lithium (i.e. inactive and not young), selected through modelling and inter-comparison of the HERMES spectra, will be the best probes of $\alpha$ across our galaxy.

\subsection{Target field selection}

\begin{figure*}
\centering
\begin{subfigure}[t]{0.49\textwidth}
    \centering
    \includegraphics[width=\columnwidth]{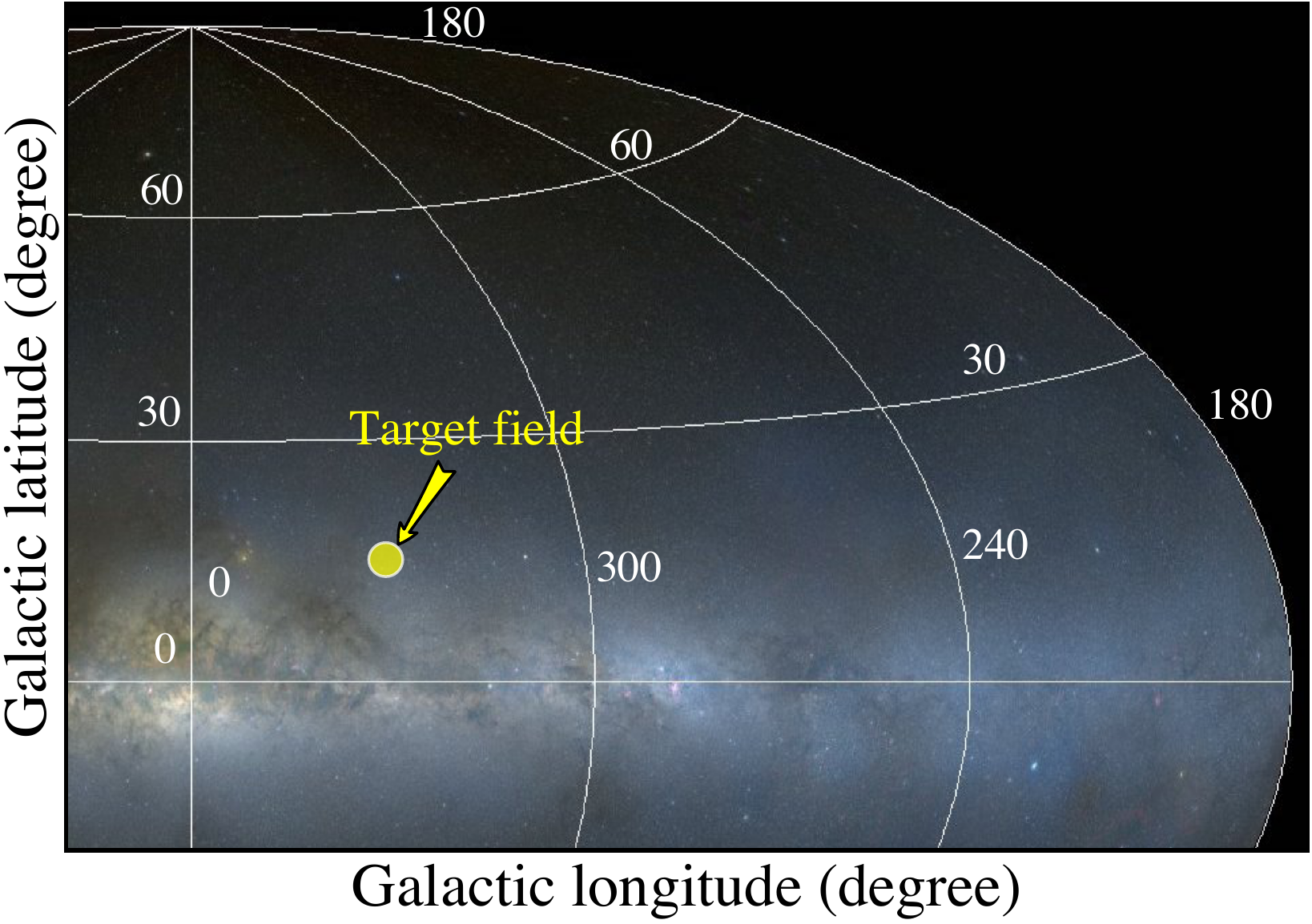} 
\end{subfigure}
\begin{subfigure}[t]{0.49\textwidth}
    \centering
    \includegraphics[width=\columnwidth]{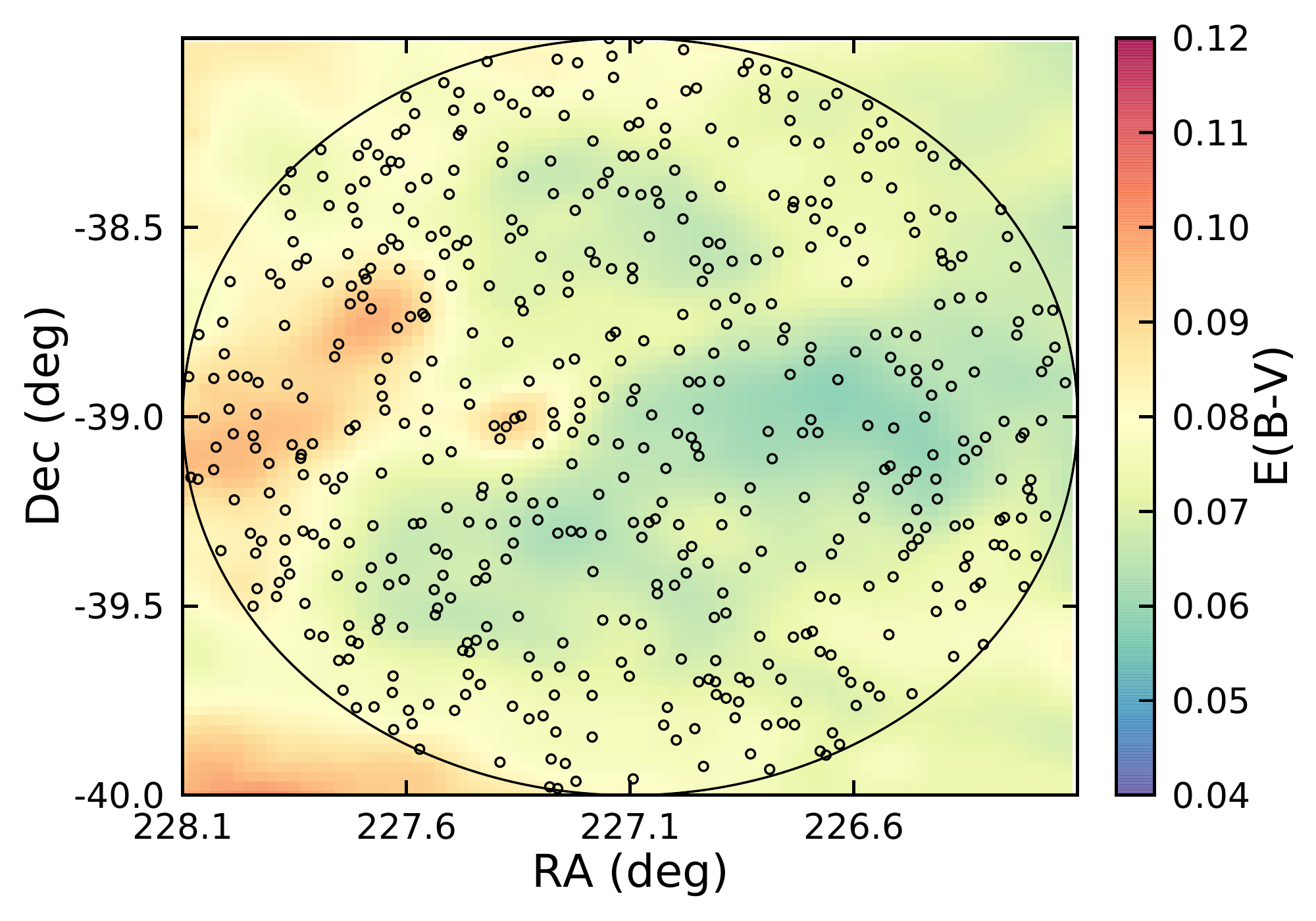} 
\end{subfigure}
\caption{Illustration of our target field. Left panel shows the Galactic location of our selected 2-degree field (background image credit: Digitized Sky Survey). Right panel shows our target field (background image credit: Skymapper) with the dust map from \citet{sch98} overlaid with our candidates observed with AAT/HERMES.} 
\label{fig:field}
\end{figure*}

We initially selected a region of sky with Galactic longitude $l$ within $\pm$30 degrees of Galactic Centre and 12 $<$ Galactic latitude $|b| <$ 30 degrees to avoid heavy dust extinction, while the region is not too far away from the direction of the Galactic Centre. Based on the dust distribution in this sky region \citep{sch98}, we then selected a 2-degree field with relatively low average reddening ($<$E(B - V)$>$ $\sim$ 0.08) as our final target field. The centre of this field is at right ascension (RA) of 227.1 degrees and declination (Dec) of $-$39.0 degrees. Fig.~\ref{fig:field} shows our selected 2-degree field with Galactic coordinates as an illustration (left panel), and the same field for AAT/HERMES observations with equatorial coordinates (right panel, overlaid with dust map and our Sun-like candidates). It demonstrates the Galactic location of our target field, avoiding the bulk of dust in the Galactic plane, and the fact that our candidates are fairly uniformly distributed in the 2-degree field with E(B $-$ V) $\lesssim$ 0.08 for most of them. 

\subsection{Target selection}

In principle, our Sun-like candidates should have similar colours and absolute magnitudes to the solar values in any given photometric system, so that they have similar \teff\ and \logg\ to those of the Sun (5772\,K and 4.44\,dex). In this work we selected our candidates in the 2-degree target field mainly based on Gaia Early Data Release 3 (EDR3; \citealp{gaia21}) with robust photometric measurements of three passbands (G, BP, and RP) and precise astrometric measurements of parallaxes. Similar to \citet{dat12} and \citet{por14}, we defined a box with small range in colours and absolute magnitudes, centered around those values of the Sun in the Gaia EDR3 photometric system \citep{cas21} to select potential targets. We also combined our catalogue of these targets with Skymapper Data Release 3 \citep{onk19} with available photometric data of three passbands ($v,g$, and $z$), in order to verify that they have solar colours in Skymapper photometric system \citep{cv18}. This also provided constraints on the range of metallicity ([Fe/H]) of our targets to exclude mainly metal-poor stars with [Fe/H] $<$ $-$0.8, because the narrow Skymapper passband $v$ provides sensitive discrimination of different metallicities. We note that pre-processing of these photometric data (e.g., zero-point correction of Gaia G passband and dereddening) was applied to reliably select Sun-like candidates with this approach. 

In addition, we made use of the spectra of a subset of GALAH Data Release 3 \citep{bud21} in the same initial sky region ($l \pm$ 30 degrees and 12 $< |b| <$ 30 degrees; $\sim$\,24800 spectra), but derived their spectroscopic parameters with the \textsc{EPIC} method, serving as a "calibration sample". The \textsc{EPIC} spectroscopic parameters and photometric/astrometric data of this "calibration sample" enable us to adjust and calibrate our photometric and astrometric criteria more reliably, eliminating potential systematic offsets, which was seen between e.g., the GALAH and \textsc{EPIC} stellar parameters (e.g., \teff, see SDST I). We note the final spectroscopic verification of solar twins and analogues from SDST will be judged based on the \textsc{EPIC} method, therefore it is self-consistent to adopt the \textsc{EPIC} spectroscopic parameters throughout the whole process of sample selection, calibration, identification and verification. 

To help explain the selection criteria, we first summarise them in four main categories here:

\begin{itemize}
    \item[] \textbf{Quality control:}
    \item $RUWE$\footnote{RUWE represents the Gaia EDR3 renormalized unit weight error parameter; see \citet{lin21}.} $ < 1.4$;
    \item Parallax\_over\_error > 2;
    \item E(B $-$ V) < 0.12;
    \item[] \textbf{Gaia photometric criteria:}
    \item $0.75 < (BP - RP)_0 < 0.87$;
    \item $0.33 < (BP - G) < 0.46$;
    \item $0.49 < (G - RP) < 0.62$;
    \item[] \textbf{Gaia astrometric criterion:}
    \item $4.1 < M_G - A_G < 5.3$;
    \item[] \textbf{Skymapper criteria:}
    \item $0.32 < {\rm Skymapper\,} (g - z)_0 < 0.52$
    \item $1.05 < {\rm Skymapper\,} (v - g)_0 < 1.45$
\end{itemize}

Below we describe the motivation and function of the selection criteria summarised above. \\
\textbf{Quality control:} As a first step to control the quality of the data used for selecting twin/analogue candidates, we selected stars from Gaia EDR3 with $RUWE < 1.4$ to ensure that their photometric measurements are robust (see \citealp{rie21} for an in-depth discussion). We excluded stars with negative Gaia parallaxes and only selected those with "parallax\_over\_error" $>$ 2. With these criteria, we retain a large fraction of stars up to distances of $\sim$4\,kpc, while the relative parallax uncertainties of our candidates are not too large. All of our candidates have reddening E(B\,$-$\,V) smaller than 0.12 to avoid heavy dust contamination, where E(B\,$-$\,V) was taken from \citet{sch98}, revised with coefficients from \citet{sf11}. 

\begin{figure}
\centering
\includegraphics[width=\columnwidth]{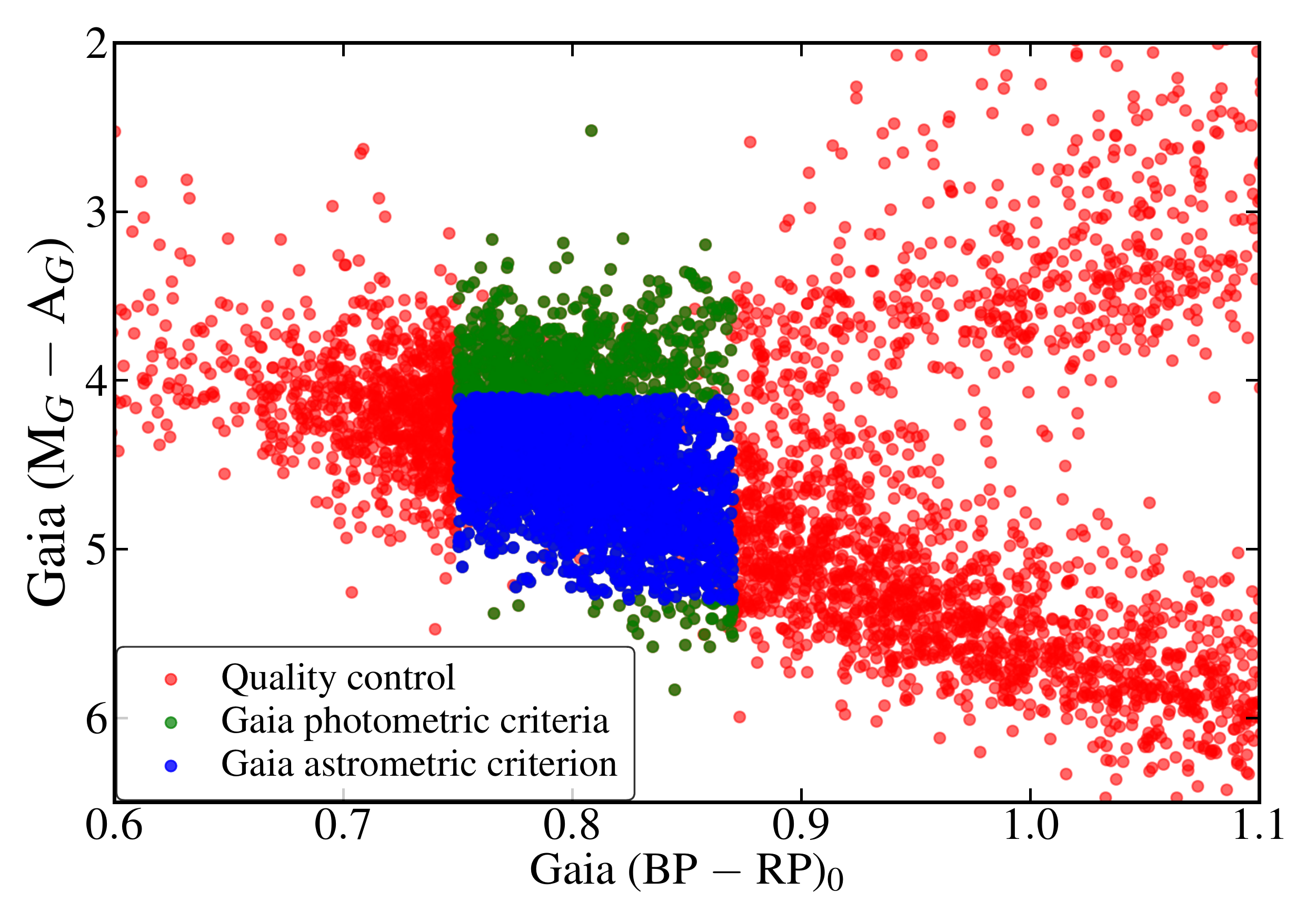}
\caption{Illustration of our selection process with Gaia EDR3 data. It shows the colour-magnitude diagram in our 2-degree target field with Gaia $(BP-RP)_0$ and $(M_G - A_G)$. The red, green, and blue points represent stars passing quality control, stars fulfilling Gaia photometric criteria, and Gaia astrometric criterion, respectively. Stars passing Skymapper criteria are not shown here because the distributions remain the same in the colour-magnitude diagram.}
\label{fig:cmd}
\end{figure}

\noindent
\textbf{Gaia photometric criteria:} We then applied pre-processing of Gaia data and the following Gaia photometric selection criteria for the second step. Zero-point corrections were applied to the Gaia G band \citep{rie21}. Reddening was taken into account and we obtained the dereddened colour index of $(BP - RP)_0$ as:
\begin{equation}
(BP - RP)_0 = BP - RP - E(BP - RP)
\end{equation}
where $E(BP - RP) = (R_{BP} - R_{RP}) \times E(B - V)$, and the extinction coefficients $R_{BP}$ and $R_{RP}$ were taken from \citet{cas21}. \\
For Gaia data, a fraction of stars (10 -- 20\%) have flux excess in passbands BP or RP relative to G, due to background contamination \citep{rie21}. Therefore additional constraints on ($BP - G$) and ($G - RP$) are necessary to remove those stars affected by the flux excess issue. By examining the colour-magnitude diagram in our target field (see Fig.~\ref{fig:cmd}), we required our candidates to fulfil the Gaia photometric criteria with: $0.75 < (BP - RP)_0 < 0.87$; $0.33 < (BP - G) < 0.46$; and $0.49 < (G - RP) < 0.62$. We note these criteria were verified with the "calibration sample", where the stars fulfilling the Gaia photometric criteria also have \textsc{EPIC}-derived \teff\ of 5772 $\pm$ 300\,K. 

\noindent
\textbf{Gaia astrometric criterion:} For the third step, we derived the absolute magnitudes in Gaia G and applied the Gaia astrometric selection criterion. The absolute magnitudes were derived as:
\begin{equation}
M_G - A_G = G - 10 + 5 \cdot log(1000/d)
\end{equation}
where the distances $d$ (in pc) were taken from \citet{bai21}, and $A_G = 2.74\times E(B - V)$ according to \citet{cv18}. \\
As demonstrated in Fig.~\ref{fig:cmd}, we required our candidates to fulfil the Gaia astrometric criterion with $4.1 < M_G - A_G < 5.3$. Again, the criterion was tested with the "calibration sample" to ensure the selected stars have \textsc{EPIC}-derived \logg\ $\pm$ 0.4\,dex around solar values. 

\noindent
\textbf{Skymapper criteria:} Finally, we combined the catalogue of our candidates with Skymapper data and applied the Skymapper selection criteria to further verify Sun-like candidates and to exclude metal-poor targets. We only selected stars with Skymapper "flags" = 0 and "g\_flags" = 0, so they have robust Skymapper photomeric measurements for at least one of the $v$, $g$ and $z$ passbands. 
We derived the dereddened colour index of e.g., Skymapper $(v-g)_0$ and $(g-z)_0$ following the method described in \citet{cas19}. 
Skymapper $(g - z)_0$ was adopted to verify our selection and to remove potential outliers due to unrealistic measurements in Gaia photometry. We note that Skymapper $(v-g)_0$ is sensitive to changes in metallicity \citep{cas19}, enabling us to effectively remove stars with lower metallicity and improve the success rate of selecting potential solar twins and analogues. By examining the results from the "calibration sample", we derived the final values for the Skymapper criteria above. The second criterion [i.e. for $(v - g)_0$] efficiently removes stars with [Fe/H] $<$ $-$0.8. 

\begin{figure*}
\centering
\begin{subfigure}[t]{0.49\textwidth}
    \centering
    \includegraphics[width=\columnwidth]{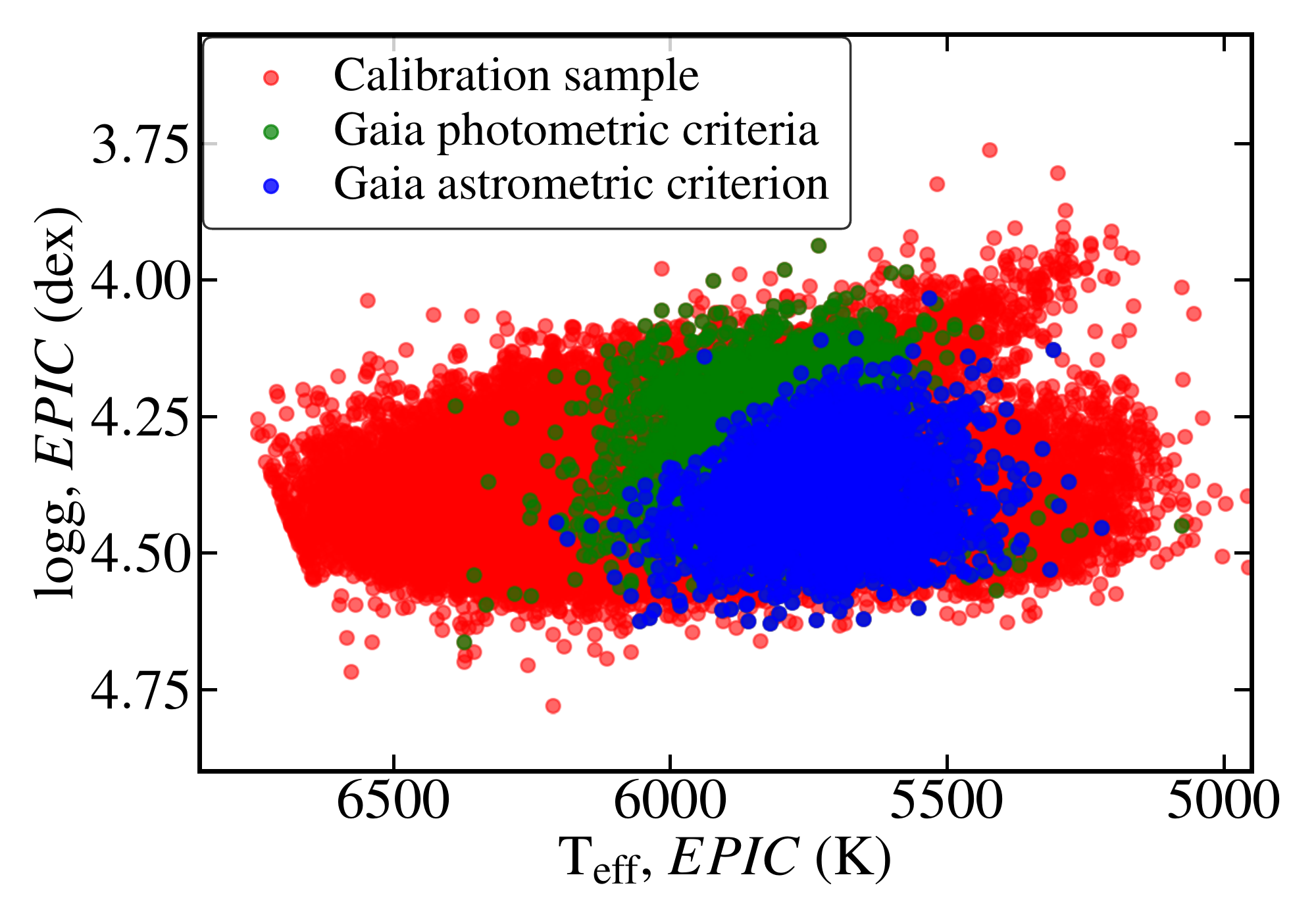} 
\end{subfigure}
\begin{subfigure}[t]{0.49\textwidth}
    \centering
    \includegraphics[width=\columnwidth]{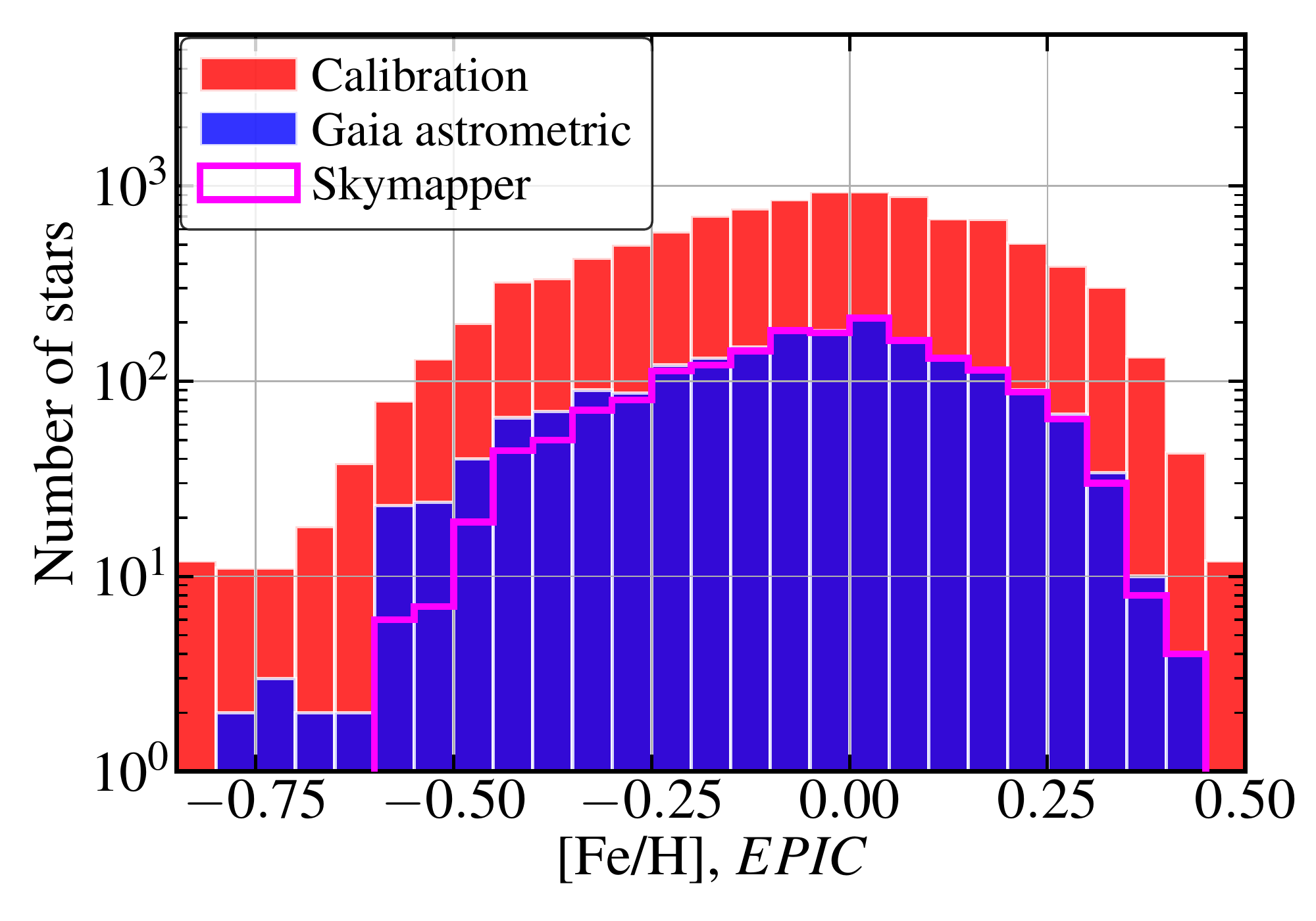} 
\end{subfigure}
\caption{Results of our target selection criteria on the "calibration sample" from the GALAH catalogue. Left panel shows the stellar parameter distributions in the \teff-\logg\ plane with Gaia phtometric (green) and astrometric criteria (blue). The distributions with Skymapper criteria are not shown since they remain almost the same in the \teff-\logg\ plane. Right panel shows the metallicity distribution with Gaia astrometric criterion (blue) and Skymapper criteria (magenta).}
\label{fig:calibration}
\end{figure*}

These target selection criteria were applied and tested on the "calibration sample", i.e. $\sim$\,24800 stars from GALAH catalogue. The results are shown in Fig.~\ref{fig:calibration} which depicts the stellar parameter distributions after applying our selection criteria (left panel: \teff\ and \logg; right panel: [Fe/H]). They demonstrate that these criteria successfully find the Sun-like stars with (\teff, \logg\ and [Fe/H]) in the range: $\pm$\,$\sim$(300\,K, 0.3\,dex and 0.6\,dex) centered around solar values. 

\begin{figure}
\centering
\includegraphics[width=\columnwidth]{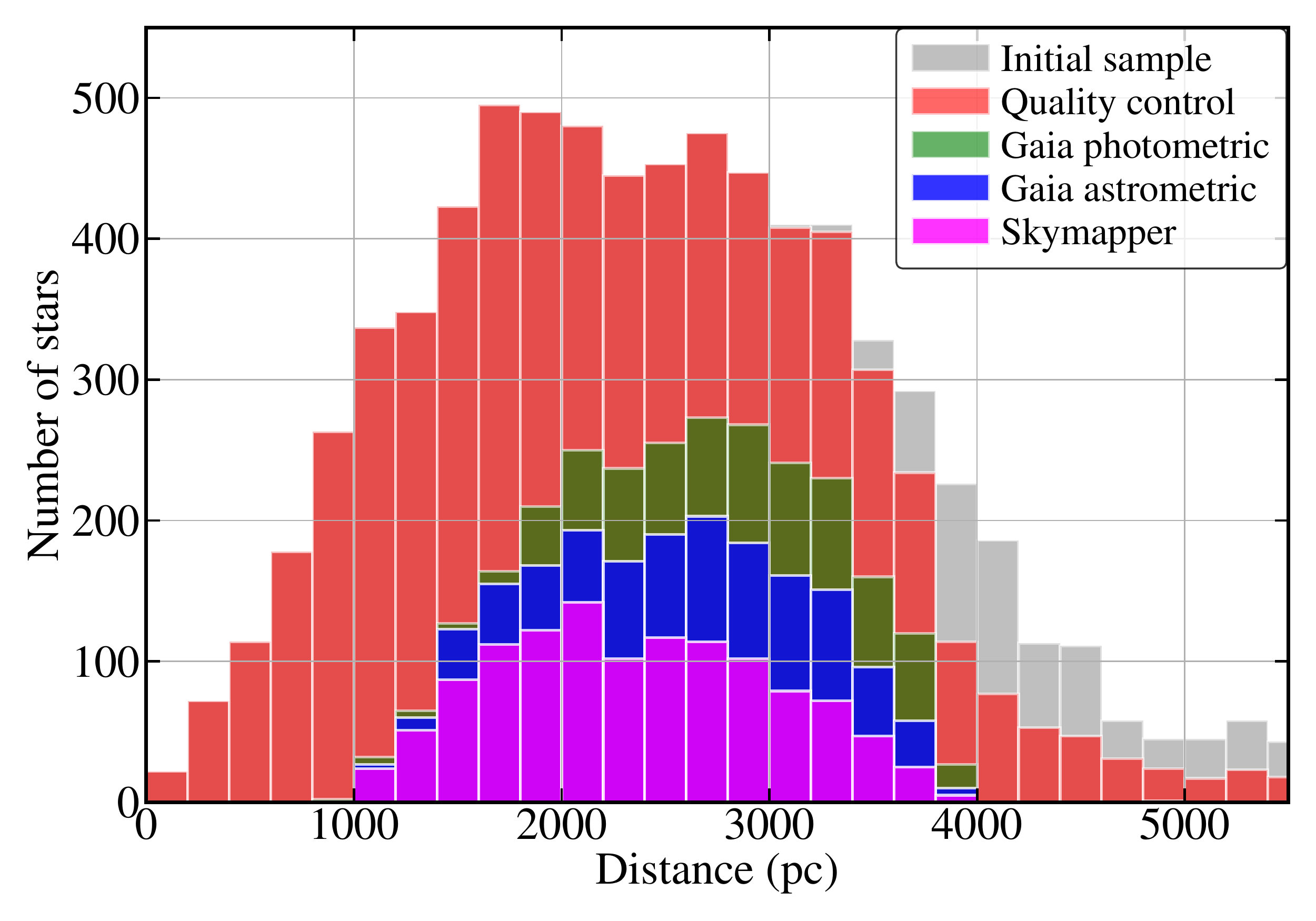}
\caption{Illustration of the stellar distance distribution of our target field, with each step of selection process. It shows the histogram of star numbers as a function of distance (in pc), with the associated selection criteria.} 
\label{fig:hist1}
\end{figure}

With these selection criteria, we have selected in total 1201 Sun-like candidates in our target field. Bearing in mind the goal of our survey is to detect solar twins and analogues 1--4\,kpc away (in several distance bins), we examined the stellar distance distribution of our selected 2-degree field after each step of selection, with the associated selection criteria in Fig.~\ref{fig:hist1}. It demonstrates the influence on the distance distribution from the above selection criteria. The quality control criteria limit us from selecting stars more distant than 4.4\,kpc, and the Gaia photometric/astrometric criteria further reduce our detectability of stars at distance $>$ 3.8\,kpc. We note this is necessary compensation to allow us to have reliable photometric and astrometric measurements with relatively low dust contamination. The distance distribution of the 1201 Sun-like candidates is centered around 2.2\,kpc, but still allows us to select enough candidates in the 3--4\,kpc distance range with our optimised observing strategy with AAT/HERMES (see Section 3.1).

\section{Observations and data reduction}

\subsection{Observing information and strategy}

We observed the selected target field with HERMES \citep{des15,she15} and the 2dF fibre positioning system \citep{lew02}, assembled on the 3.9\,m AAT at Siding Spring Observatory. The observations were taken on 7\,-\,15 June and 7\,-\,11 July, 2021 during dark time, under AAT program A/2021A/005. HERMES is a multi-fibre fed spectrograph with 400 fibres and up to 392 stars can be observed simultaneously\footnote{25--28 fibres were broken or disabled during our observing run.}. It has four bands (blue, green, red, and infrared) that cover four wavelength ranges: 4713 -- 4903, 5647 -- 5873, 6478 -- 6737, and 7585 -- 7887 \AA, respectively. The spectral resolving power is $\sim$ 28,000 although it varies slightly from fibre to fibre, and across each wavelength range \citep{kos17}. 

\begin{table}
\centering
\caption{Journal of AAT/HERMES observing and conditions. The observing date, status of fibre configuration (Config) and 2dF fibre positioning plate (P), exposure time ($T_{\rm exp}$), airmass and approximate seeing are listed in column\,1--6, respectively.} 
\label{tab:obs}
\begin{tabular}{@{}cccccc@{}}
\hline
Observing & Config & P & $T_{\rm exp}$ & Airmass & Seeing \\
date & & & (sec) & & (") \\
\hline 
07-Jun-21 & Config\,B & 0 & 900 & 1.52 & 1.5 \\
07-Jun-21 & Config\,1 & 1 & 3900 & 1.40 & 1.6 \\
07-Jun-21 & Config\,1 & 0 & 3900 & 1.14 & 1.8 \\
07-Jun-21 & Config\,1 & 1 & 3900 & 1.03 & 1.6 \\
07-Jun-21 & Config\,1 & 0 & 3900 & 1.02 & 1.7 \\
13-Jun-21 & Config\,2 & 0 & 4200 & 1.17 & 2.3 \\
13-Jun-21 & Config\,2 & 0 & 3600 & 1.01 & 1.3 \\
13-Jun-21 & Config\,2 & 1 & 3900 & 1.04 & 1.6 \\
14-Jun-21 & Config\,B & 1 & 1080 & 1.38 & 1.4 \\
14-Jun-21 & Config\,2 & 0 & 3960 & 1.30 & 1.5 \\
14-Jun-21 & Config\,2 & 1 & 3960 & 1.09 & 1.6 \\
14-Jun-21 & Config\,3 & 0 & 4200 & 1.01 & 1.1 \\
14-Jun-21 & Config\,3 & 1 & 4320 & 1.03 & 1.3 \\
15-Jun-21 & Config\,B & 1 & 1020 & 1.36 & 1.4 \\
15-Jun-21 & Config\,3 & 0 & 3780 & 1.28 & 1.5 \\
15-Jun-21 & Config\,3 & 1 & 3780 & 1.09 & 1.4 \\
15-Jun-21 & Config\,3 & 0 & 3900 & 1.02 & 1.4 \\
15-Jun-21 & Config\,1 & 1 & 3900 & 1.02 & 1.6 \\
15-Jun-21 & Config\,4 & 0 & 3960 & 1.12 & 1.6 \\
07-Jul-21 & Config\,4 & 1 & 4500 & 1.06 & 2.0 \\
07-Jul-21 & Config\,4 & 1 & 4500 & 1.06 & 1.5 \\
10-Jul-21 & Config\,B & 1 & 1080 & 1.07 & 1.8 \\
10-Jul-21 & Config\,4 & 1 & 4400 & 1.02 & 1.6 \\
10-Jul-21 & Config\,4 & 0 & 4200 & 1.14 & 1.6 \\
11-Jul-21 & Config\,B & 0 & 1020 & 1.07 & 1.8 \\
11-Jul-21 & Config\,4 & 1 & 4250 & 1.05 & 1.5 \\
11-Jul-21 & Config\,4 & 0 & 4500 & 1.01 & 1.5 \\
11-Jul-21 & Config\,4 & 1 & 3900 & 1.08 & 2.0 \\
\hline
\end{tabular}
\end{table}

To achieve our goal of spectroscopically verifying solar twin and analogue candidates with the EPIC method (Section 2.3), it is necessary to obtain HERMES spectra of our targets with SNR above a threshold of $\gtrsim$\,20-25 per pixel in the red band as discussed in SDST I. This SNR threshold, which is fairly low compared with the requirements of other methods, is required for \textsc{EPIC} to work effectively and accurately determine spectroscopic parameters of Sun-like stars. In addition, we aim to observe enough candidates in different magnitude bins, so that we can identify enough solar twins/analogues for follow-up at all distances 1--4\,kpc. This is particularly important and challenging for the faintest (most distant) candidates, which sets our optimised observing strategy as described below. 

The survey strategy was designed to ensure that a similar SNR could be achieved for all targets. We therefore separated our sample stars into three magnitude bins: 15.4 -- 16.2\,mag, 16.2 -- 16.8\,mag, and 16.8 -- 17.4\,mag, which effectively corresponds to three distance bins. Targets in the faintest bin were observed in all 28 HERMES exposures, while those in the middle bin were observed in half the exposures, and the brightest targets were observed only (approx.) one quarter of the time. In more detail: \\
1. 547 candidates were selected, with 184, 183 and 180 stars in the bright, medium and faint bins respectively, for observations in four different fibre configurations (hereafter `Config\,1--4'). We refer them as our "main targets/candidates" hereafter. \\
2. All configurations included the fibres allocated to the same 180 stars in the faint bin. Two sets of candidates ($\approx$\,91--93 stars each) in the medium bin were allocated to the corresponding fibres for `Config\,1-2' and `Config\,3-4'. Four sets of candidates ($\approx$\,45--47 stars each) in the bright bin were allocated to the remaining fibres to each of the four configurations. \\
3. 330 brighter candidates with magnitude 13.4 $<$ Gaia G $<$ 15.4\,mag were observed during twilight, in a single configuration (hereafter `Config\,B'). We note these were not observed during dark time and, throughout our two runs, only 5 valid exposures were obtained for them, resulting in many with SNR $<$ 20. \\ 
4. 40 fibres were allocated to sky positions for each fibre configuration, so that we could reliably combine the sky spectra from each exposure, which is essential for further correction (see Section 3.2). 

To summarise, we have obtained 23 effective exposures\footnote{Poor weather conditions during additional 5 exposures meant they did not contribute meaningfully to the total SNR, so they were not analysed or included any further.} with an average exposure time of 4056\,sec for the main candidates, and 5 effective exposures with an average exposure time of 1020\,sec for the bright, twilight candidates (G $<$ 15.4\,mag). Following the above observing strategy, we obtained AAT/HERMES spectra of 877 sample stars in total, including 547 main candidates and 330 bright candidates. The journal of AAT/HERMES observing and conditions is presented in Table~\ref{tab:obs}. We expected to build up the necessary SNR ($\geq$ 25 in the red band) for our main targets by combining the spectra of all exposures.

\subsection{Data reduction}

The data reduction was conducted using the \textit{2dfdr} reduction software (version 7.3; \citealp{aao15}) with minor modifications. We made use of \textit{2dfdr} pipeline to apply bias subtraction, fibre tracing, wavelength calibration, optimal extraction (minimising the effect of scattered light and cross-talk; \citealp{sb10}), flat fielding, and barycentric velocity correction. The final extracted spectra were then rebinned to the same wavelength grid for the four HERMES bands (blue, green, red, and infrared) with a pixel size of 0.0452, 0.0547, 0.0632, and 0.0737 \AA, respectively. 

We applied custom sky subtraction and telluric correction for each exposure. Note that the spectral lines of interest (see details of spectral line selection in SDST I) should not be directly affected by these two steps, therefore it is not crucial to make them perfectly. However, given the fact that we will subsequently combine the spectra of each star from multiple exposures with different scaling factors, the final data product can be improved after reasonable sky subtraction and telluric corrections. In addition, with the combined spectrum of a single object, we need to set the continuum, globally in each band and locally around each spectral line of interest for \textsc{EPIC} to work well (SDST I), which can also benefit from these two steps. We describe the details of sky subtraction and telluric correction below. 

\begin{figure*}
\centering
\includegraphics[width=0.98\textwidth]{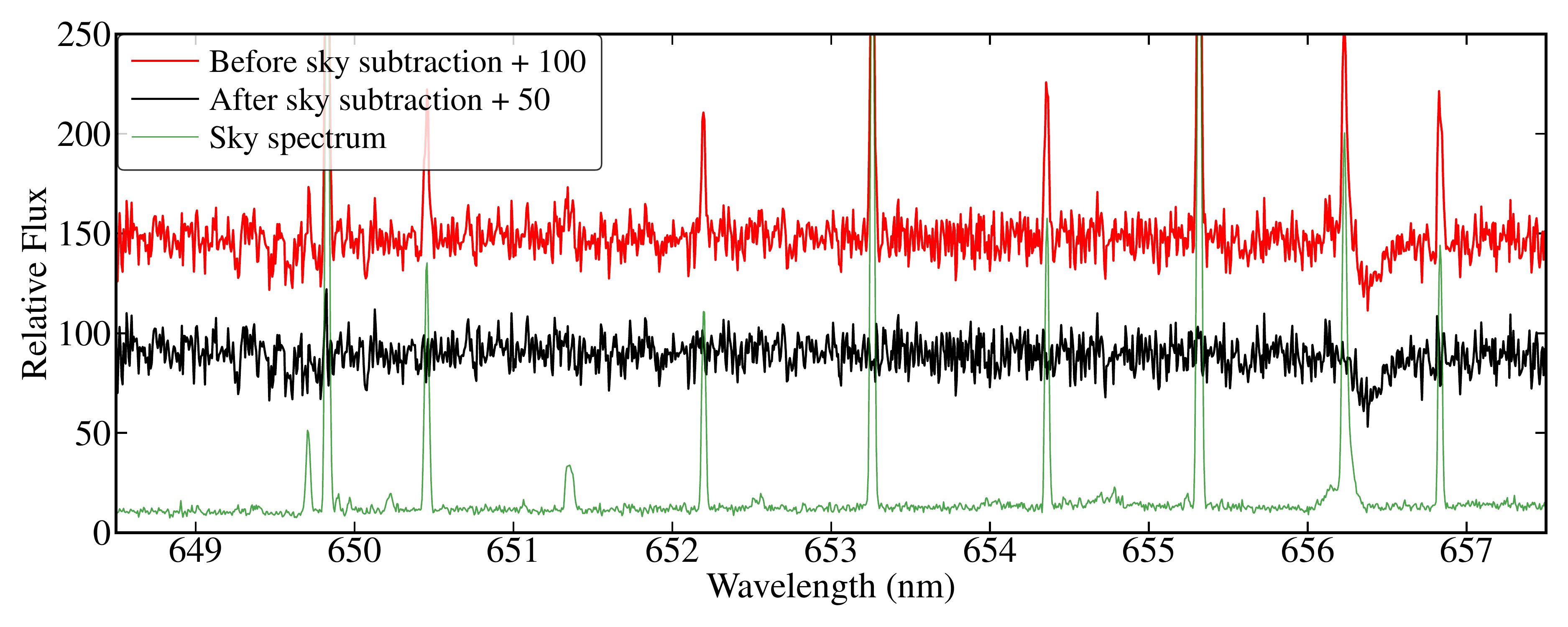}
\caption{Illustration of custom sky subtraction results. It shows an example of sky subtraction for a typical target from a single exposure. The median sky spectrum, spectra before and after sky subtraction (shifted by $+$100 and $+$50 flux counts) are shown in green, red and black, respectively.} 
\label{fig:sky_sub}
\end{figure*}

\noindent
\textbf{Sky subtraction:} Our own sky subtraction algorithm was applied to take into account carefully the effect of different fibre throughput. For each exposure, we divided each of the 40 sky spectra by the relative throughput of the corresponding fibre, derived from its corresponding flat-field exposure, to scale them to equivalent flux levels initially. The median sky spectrum for each exposure was then obtained using these scaled sky spectra. Similarly we divided each object spectrum by its relative fibre throughput for each exposure. The median sky spectrum was then subtracted from each object spectrum. An example of a stellar spectrum before and after sky subtraction is shown in Fig.~\ref{fig:sky_sub}, demonstrating that we efficiently remove strong sky features. 

\begin{figure*}
\centering
\begin{subfigure}[t]{0.98\textwidth}
    \centering
    \includegraphics[width=\columnwidth]{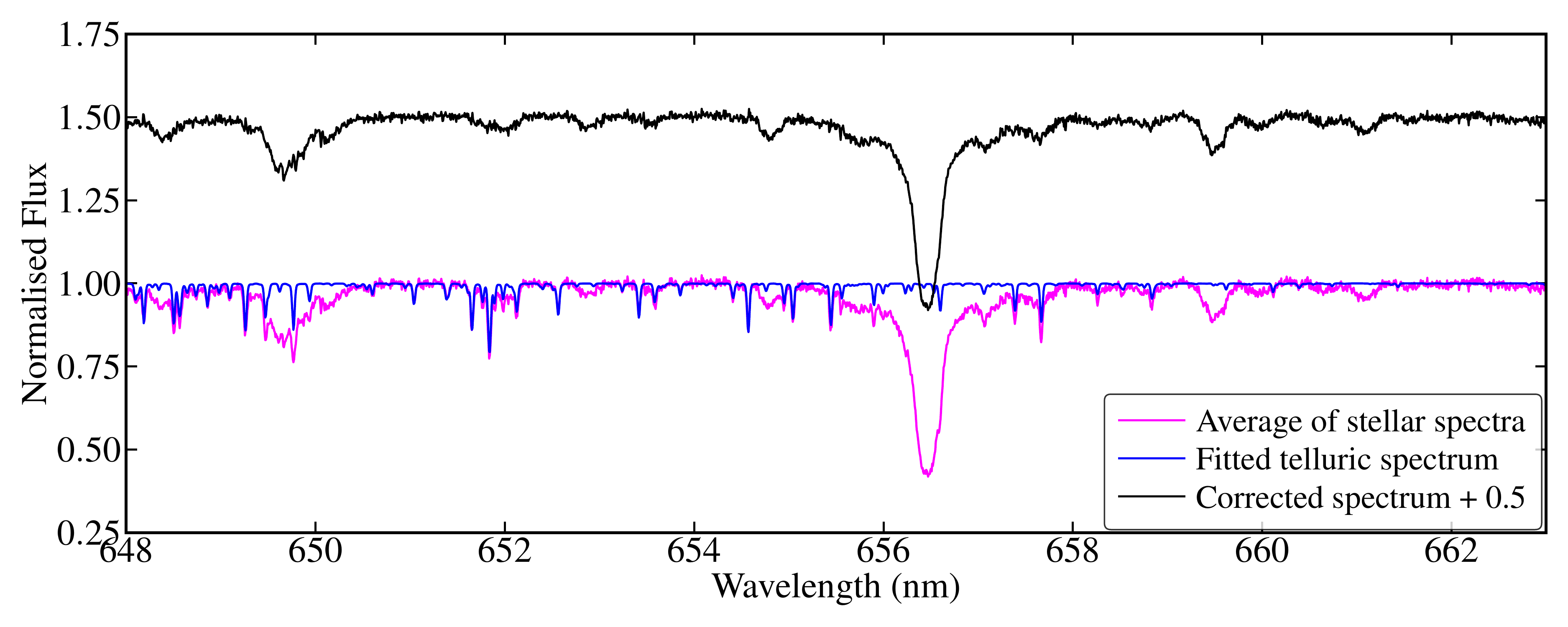}
\end{subfigure}
\begin{subfigure}[t]{0.98\textwidth}
    \centering
    \includegraphics[width=\columnwidth]{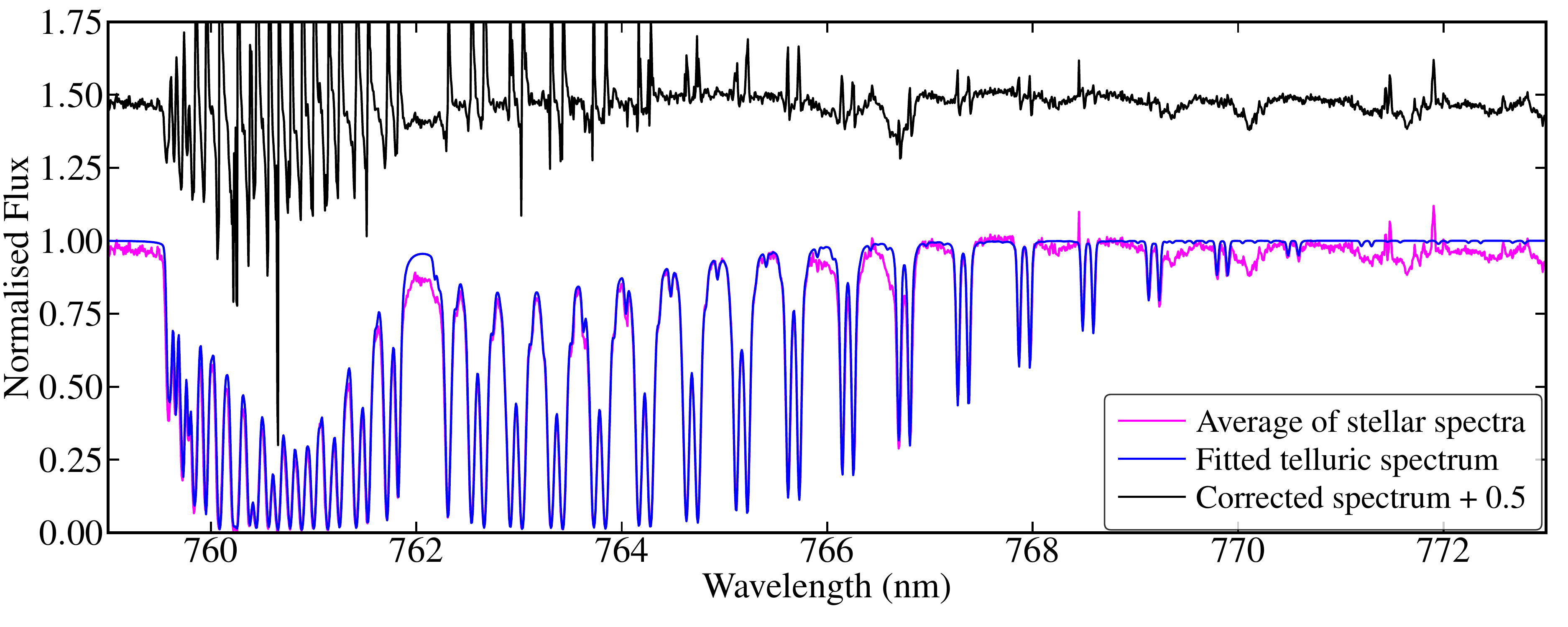} 
\end{subfigure}
\caption{Example of telluric correction in a portion of the red band (upper panel) and infrared band (lower panel) for one exposure. The original spectrum, the transmission spectrum calculated from \textit{Molecfit}, and the telluric corrected spectrum are shown in magenta, blue, and black, respectively.}
\label{fig:telluric}
\end{figure*}

\noindent
\textbf{Telluric correction:} The \textit{Molecfit} software \citep{sme15,kau15} was applied for the telluric correction of our reduced spectra for the red and infrared bands, because the telluric absorption lines are only strong enough to be detectable and corrected in these two bands. We note that, given their low SNR, it would be difficult, time-consuming and ultimately ineffective to run \textit{Molecfit} on each star spectrum in each exposure. Instead, we applied \textit{Molecfit} on the average spectrum derived from all stellar spectra in a single exposure. Within the 2-degree field, the airmass differences between object spectra should be minor. Therefore we can use the transmission spectrum calculated by \textit{Molecfit} for all the targets from the same exposure. The process of telluric correction is described below: \\
1. A weighted mean spectrum was obtained using all the sky-subtracted stellar spectra of a single exposure, after scaling individual spectra (and corresponding uncertainty arrays from \textit{2dfdr}) to the same median flux level in each of the red and infrared bands. The scaled flux and inverse variance spectra (the latter derived from the scaled uncertainty spectra) can then be used to generate the weighted mean spectrum. This left us an average spectrum of each exposure with telluric features and some intrinsic stellar features which appear broadened due to radial velocity differences between objects. \\ 
2. We run \textit{Molecfit} on the weighted mean spectrum of each exposure to find the best-fit molecular models and the normalised transmission spectrum. The strongest intrinsic stellar features and bad pixels were identified by eye and excluded before the fitting process. \\ 
3. Each object spectrum was divided by the corresponding transmission spectrum for each exposure. This provided us with the reduced spectra with telluric correction applied. An example of reduced spectra before and after telluric correction is shown in Fig.~\ref{fig:telluric}, demonstrating that we reasonably removed a large number of telluric features. We note the overcorrection in a portion of the infrared band has no impact on the \textsc{EPIC} analysis, because none of our adopted spectral lines are located in the affected regions (SDST III).

The reduced spectra of each target were then combined using the weighting scheme of the \textit{2dfdr} pipeline (as documented in \citealp{aao15}).

\section{Survey verification}

\subsection{Final data description}

\begin{figure*}
\centering
\includegraphics[width=0.98\textwidth]{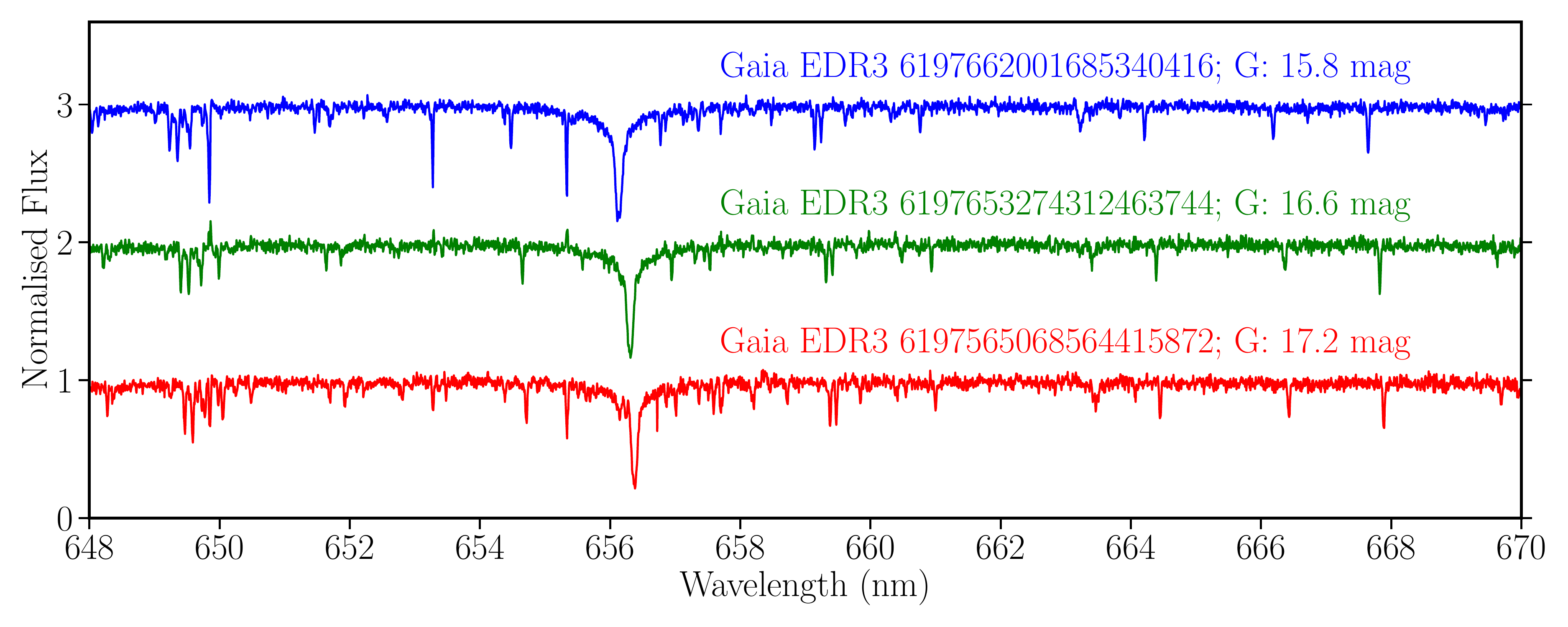}
\caption{Examples of our reduced spectra in the HERMES red band for three typical sample stars with continuum normalised. The blue, green, and red spectra represent stars with magnitude of 15.8, 16.6, and 17.2, respectively.}
\label{fig:spec}
\end{figure*}

\begin{figure*}
\centering
\includegraphics[width=0.98\textwidth]{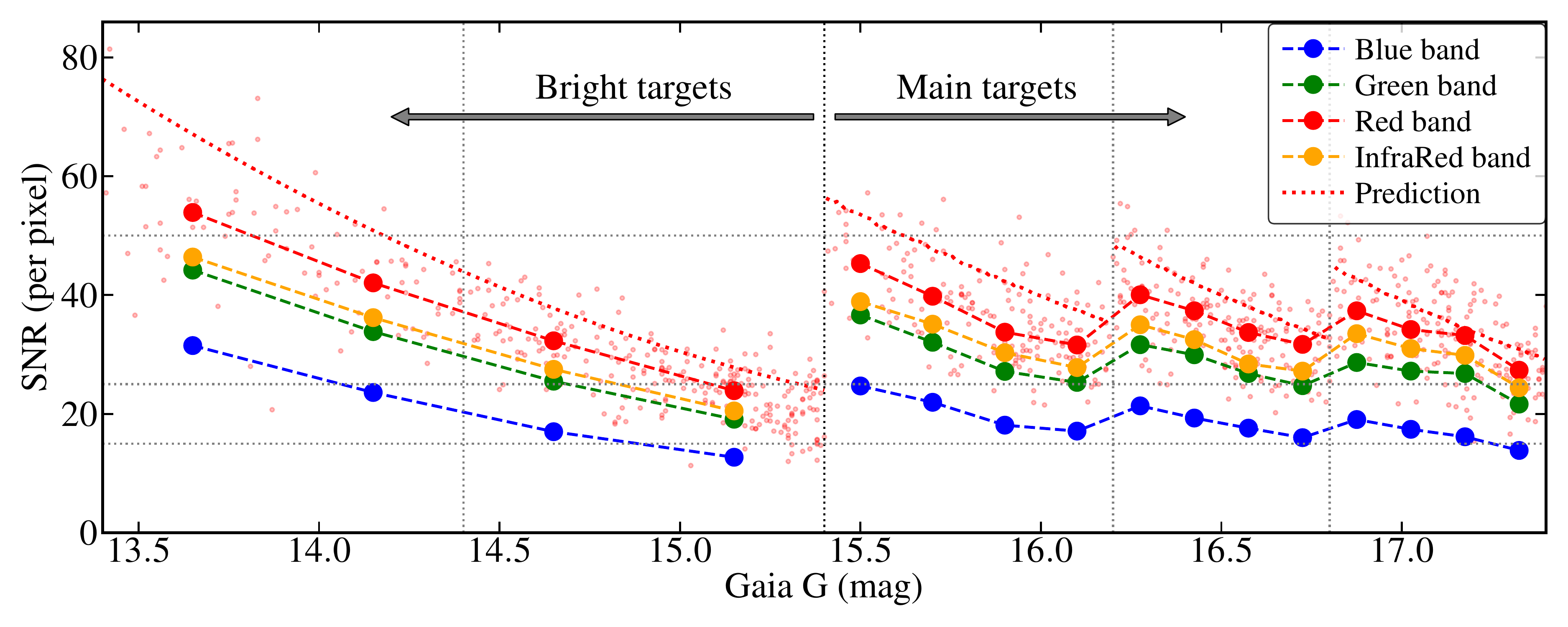}
\caption{SNR of our reduced spectra in the four HERMES bands as a function of Gaia G magnitude of our main and bright targets. The small red dots represent SNR in the red band for each target. The big dots in blue, green, red, and orange, connected with dashed lines, represent the obtained average SNR of each magnitude bin in the blue, green, red, and infrared bands, respectively. The red dotted line shows our expectations based on the basic exposure time calculations for the red band (which did not include realistic degradation or variation in seeing or fibre throughput).}
\label{fig:snr}
\end{figure*}

The final reduced 1-dimension spectra of all 877 targets are available at \url{https://datacentral.org.au}. Similar to the data structure of GALAH spectra \citep{kos17}, the combined spectrum of each target was stored as four FITS files (1--4) corresponding to four HERMES bands (blue, green, red and infrared). Each FITS file includes the relative flux and the corresponding uncertainty array. Fig.~\ref{fig:spec} shows the reduced spectra of three typical sample stars spanning the magnitude range 15.8--17.2\,mag (cf.\ our full magnitude range for the main targets of 15.4--17.4\,mag). The SNR of the reduced spectra for the four HERMES bands were estimated using the selected wavelength range of: 4730 -- 4840\,\AA\ (blue), 5670 -- 5860\,\AA\ (green), 6500 -- 6540 and 6590 -- 6720\,\AA\ (red), and 7670 -- 7870\,\AA\ (infrared), in order to avoid strong stellar features. The 75$^{\rm {th}}$ percentile flux and the corresponding uncertainty (from the produced error array) were then adopted for SNR measurements of each HERMES band. 
Fig.~\ref{fig:snr} shows the SNR in the four HERMES bands of our target spectra as a function of Gaia magnitudes. This illustrates that we achieved similar SNR across the magnitude range, enabling us to measure stellar parameters in stars at all distances up to 4\,kpc, with low enough uncertainties, to select a sample of at least 30 for future follow-up at higher resolution and SNR. In addition, we note the obtained spectral SNR at a given magnitude shows significant scatter, mainly due to different fibre throughputs among the \textit{2df} fibres and the fact that some targets were always allocated to the fibres with much lower throughput. Nevertheless, we managed to build up the required SNR in the red band for about 70\% of our targets, which is sufficient to satisfy our survey goals. 

We emphasize that it is extremely challenging to obtain the spectra of reasonable quality for spectroscopic analysis with stellar magnitude $>$ 16\,mag. Indeed, to our knowledge, no previous HERMES observations have been conducted on targets fainter than G $\sim$ 15\,mag, so our observations, with very long total exposure times, extend at least 2 magnitudes deeper. Therefore, spectroscopic verification of these faint candidates provide us with a completely new knowledge of distant Sun-like stars in several distance bins from 1--4\,kpc in the inner region of Milky Way. In comparison, the most distant solar twin known so far is only 800--900\,pc away, located in the solar neighbourhood \citep{one11,liu16}.

\subsection{Stellar properties}

The stellar properties (e.g., \teff, \logg, mass) of our candidates were derived using the Gaia EDR3 photometric and astrometric data for initial verification. The photometric \teff\ of our targets were derived following the method presented in \citet{cas21} with fixed \logg\ = 4.0\,dex and [Fe/H] = 0.0\,dex as initial input. The bolometric magnitude $M_{\rm bol}$ can be determined with absolute Gaia G magnitude as:
\begin{equation}
M_{\rm bol} = M_G - A_G + BC_G
\end{equation}
where the bolometric corrections ($BC$) in Gaia G for our targets were initially derived using the photometric \teff, fixed \logg\ of 4.0\,dex and [Fe/H] of 0.0\,dex, following the method described in \citet{cv18}.

The stellar masses were determined using the Yonsei-Yale isochrones \citep{dem04}, adopting the photometric \teff\ and $M_{\rm bol,G}$ as initial input. The \logg\ values of our targets can then be derived as:
\begin{equation}
\log g = \log M_{\rm star} + 4 \times \log \frac{T_{\rm eff,star}}{T_{\rm eff,\odot}} + 0.4 \times (M_{\rm bol,star} - M_{\rm bol,\odot}) + \log g_{\odot}
\end{equation}

The above processes were iterated using the derived \logg\ as input until they converged, adopting the [Fe/H] values determined spectroscopically (see SDST III). We note metallicities do not significantly change the derived stellar masses and \logg\ values. The stellar information and derived properties of a small portion of our observed candidates are listed in Table~\ref{tab:param}. The full table is available online. 

\begin{figure}
\centering
\includegraphics[width=\columnwidth]{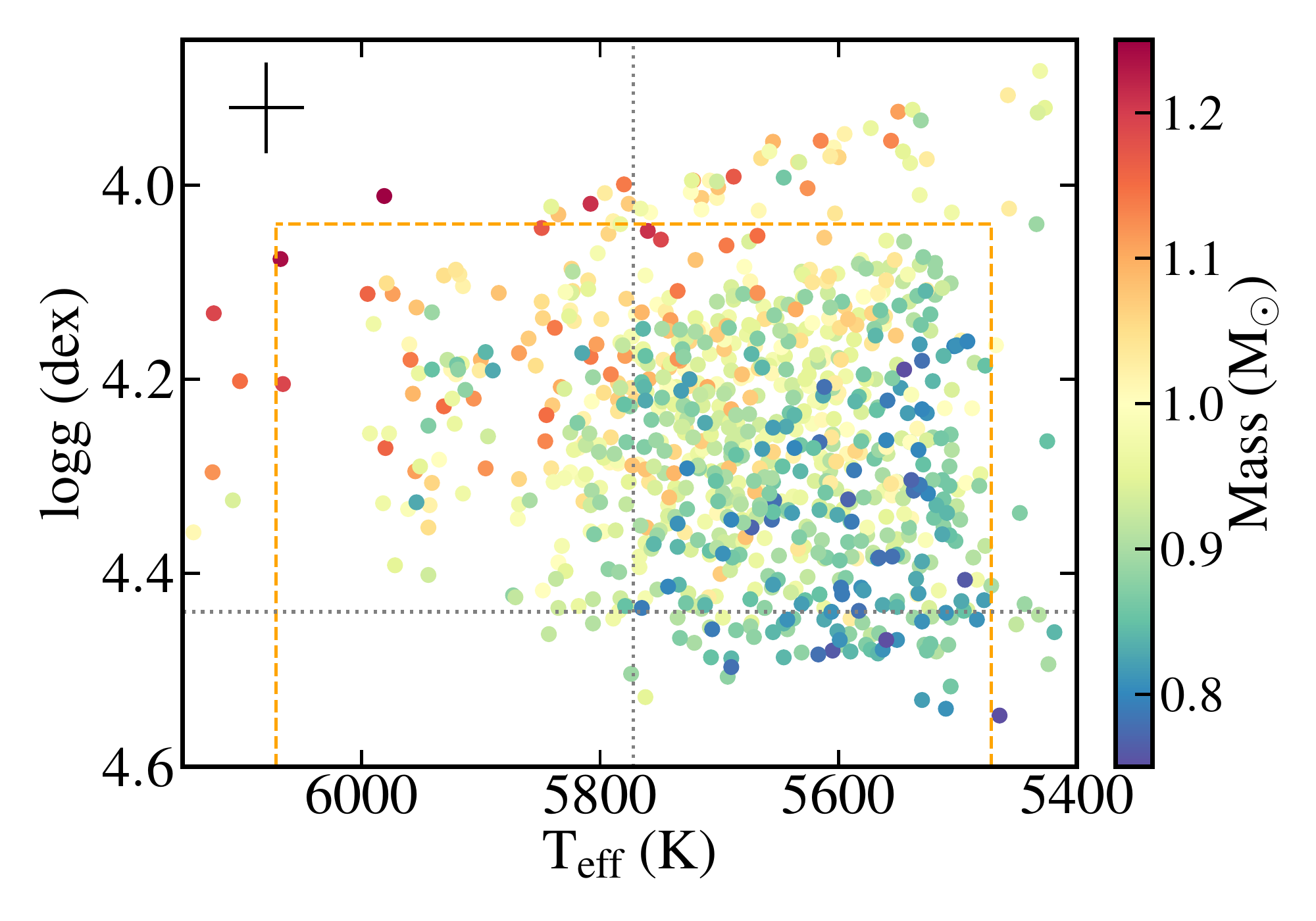}
\caption{Illustration of the stellar properties of our candidates derived based on the Gaia EDR3 photometric and astrometric data. It shows their effective temperature as a function of surface gravity, colour-coded with stellar masses. The grey dotted lines mark out the solar value of \teff\ and \logg; the orange dashed lines indicate our definition of solar analogues in the \teff-\logg\ space. We note none of our targets have \logg\ $>$ 4.6\,dex. A typical error bar (60\,K in \teff and 0.09\,dex in \logg) is plotted in black.}
\label{fig:para}
\end{figure}

The stellar properties (\teff, \logg, and mass) of our targets, from the photometric and astrometric data, are shown in Fig.~\ref{fig:para}. The results demonstrate that the majority of these candidates ($\gtrsim$ 92\%) are within our definition of solar analogues (as stated in Section 1) in terms of their \teff\ and \logg\ values, and $\approx$ 63\% of them have stellar masses between 0.9-1.1\,$M_{\odot}$. However, we do require accurate stellar metallicity [Fe/H] from spectroscopy to verify their likeness to the Sun, and this will be computed with the EPIC algorithm (SDST I) and presented in SDST III. Nevertheless, these stellar properties still provide us an independent estimation about how many of our candidates are potential solar analogues. An in-depth discussion about our selection success rate of solar analogues is presented in Section 4.3.

In addition, we note that the distribution of \logg\ seems slightly broad and skewed towards a peak value of $\sim$ 4.3--4.35\,dex. This is mainly due to the systematic difference in spectroscopic \logg\ with \textsc{EPIC} (for the "calibration sample" used in this paper) and photometric/astrometric \logg\ ($\sim$ 0.1\,dex). We note such a systematic difference was reported and discussed in previous studies (see e.g., \citealp{ben14} and \citealp{cas20}). Meanwhile we would naturally expect many fewer stars with \logg $>$ 4.6 at solar temperature, similar to e.g., the GALAH survey (see Fig.~5 and Fig.~24 in \citealp{bud21}), which also leads to the slightly skewed \logg\ distribution. 

\begin{table*}
\centering
\begin{minipage}{190mm}
\caption{Stellar information and properties of our sample stars.} 
\label{tab:param} 
\begin{tabular}{@{}lccccccccccc@{}}
\hline
Gaia EDR3 ID & Gaia G & Parallax$^a$ & $\sigma$Parallax & Distance$^b$ & $(BP - RP)_0$ & $M_G - A_G$ & \teff & $\sigma$\teff & \logg & $\sigma$\logg & Mass \\
 & (mag) & ($mas$) & ($mas$) & (pc) & (dex) & (mag) & (K) & (K) & (dex) & (dex) & (\msun) \\
\hline 
 6005408132963970176 & 17.036 & 0.299 & 0.087 & 2960 & 0.851 & 4.462 & 5551 & 64 & 4.185 & 0.049 & 0.967 \\
 6197849983813131648 & 16.046 & 0.551 & 0.049 & 1727 & 0.841 & 4.638 & 5579 & 55 & 4.260 & 0.042 & 0.940 \\
 6005435208437589632 & 15.579 & 0.567 & 0.043 & 1632 & 0.772 & 4.272 & 5774 & 58 & 4.209 & 0.053 & 1.042 \\
 6197709280686007808 & 16.313 & 0.458 & 0.057 & 2073 & 0.765 & 4.475 & 5792 & 60 & 4.286 & 0.061 & 1.007 \\
 6005389784863513600 & 15.740 & 0.500 & 0.051 & 1893 & 0.802 & 4.112 & 5686 & 56 & 4.114 & 0.060 & 0.981 \\
 6197686397100825856 & 15.643 & 0.772 & 0.049 & 1237 & 0.808 & 4.945 & 5670 & 56 & 4.417 & 0.046 & 0.959 \\
 6005400298943595776 & 15.728 & 0.544 & 0.040 & 1712 & 0.813 & 4.328 & 5655 & 56 & 4.180 & 0.079 & 0.979 \\
 6197677223051531392 & 16.217 & 0.469 & 0.058 & 2005 & 0.843 & 4.502 & 5579 & 55 & 4.216 & 0.046 & 0.976 \\
 6197723157722392576 & 16.046 & 0.481 & 0.051 & 1985 & 0.783 & 4.321 & 5744 & 62 & 4.213 & 0.054 & 1.006 \\
 6197557307562451840 & 16.558 & 0.386 & 0.065 & 2422 & 0.824 & 4.408 & 5625 & 56 & 4.195 & 0.068 & 0.960 \\
 6197726769792974592 & 16.527 & 0.389 & 0.058 & 2428 & 0.814 & 4.349 & 5645 & 59 & 4.176 & 0.073 & 0.955 \\
 6005470736406333696 & 15.955 & 0.557 & 0.049 & 1683 & 0.785 & 4.555 & 5725 & 60 & 4.284 & 0.053 & 0.987 \\
 6005529079242610176 & 15.722 & 0.607 & 0.040 & 1517 & 0.791 & 4.511 & 5692 & 61 & 4.240 & 0.051 & 0.970 \\
 6197533221385791104 & 15.749 & 0.621 & 0.040 & 1499 & 0.804 & 4.639 & 5684 & 56 & 4.307 & 0.056 & 0.961 \\
 6197681792894913792 & 15.556 & 0.584 & 0.046 & 1610 & 0.816 & 4.320 & 5658 & 53 & 4.186 & 0.050 & 0.954 \\
 6197694780876192640 & 16.107 & 0.533 & 0.047 & 1790 & 0.800 & 4.570 & 5679 & 62 & 4.267 & 0.103 & 0.971 \\
 6197563045638807168 & 15.726 & 0.477 & 0.049 & 1901 & 0.775 & 4.107 & 5773 & 55 & 4.156 & 0.041 & 1.011 \\
 6197651483308551680 & 16.560 & 0.348 & 0.061 & 2638 & 0.786 & 4.247 & 5747 & 59 & 4.198 & 0.052 & 0.999 \\
 6197741505821741952 & 16.745 & 0.460 & 0.089 & 2269 & 0.805 & 4.740 & 5681 & 60 & 4.345 & 0.062 & 0.954 \\
 6197793255886887552 & 16.613 & 0.491 & 0.072 & 1885 & 0.846 & 4.987 & 5554 & 60 & 4.375 & 0.072 & 0.944 \\
 \hline
\end{tabular}
\\
$^a$ Parallaxes and their uncertainties are taken from Gaia EDR3 \citep{gaia21}. \\
$^b$ Distances are taken from \citet{bai21}. \\
This table is published in its entirety in the electronic edition of the paper. A portion is shown here for guidance regarding its content.
\end{minipage}
\end{table*}

\subsection{Selection success rate of solar analogues}

In this work we simply defined the "selection success rate" as the fraction of stars belonging to our definition of solar analogues (as stated in Section 1), as determined from the Gaia photometric and astrometric data, out of stars fulfilling our selection criteria (as listed in Section 2.3). We estimated and examined the selection success rate of solar analogues in our target field and how it changes with increasing Gaia G magnitudes (fainter and more distant). A realistic estimation of the selection success rate is important because it needed to be incorporated into the planning of the observations, and it is essential for us to understand the final spectroscopic measurements made in SDST III, especially whether the number of solar analogues identified is reasonable. 

In Section 4.3.1--4.3.3, the expected selection success rate of solar analogues is assessed using a simulated data set, another set from a theoretical stellar population model, and is also estimated based on the results from both the photometric/astrometric parameters from this paper and the spectroscopic parameters from SDST III. The selection success rate estimated from different methods are compared and discussed in Section 4.3.3.

\subsubsection{Success rate based on simulated data set}

The overall idea of our calculation in this subsection is to estimate how efficiently we can select faint solar analogues (15.4--17.4\,mag) in our 2-degree target field with the selection criteria specified in Section 2.3. However, we do not have a set of solar-like stellar spectra in the same magnitude range with which to test the effectiveness of our selection criteria. Therefore we had to simulate a data set with fainter magnitude range and corresponding realistic uncertainties using a subset of GALAH spectra with brighter stars to represent our preliminary expectation before our AAT observations. For this purpose, a bright subset of stars (11.0--15.5\,mag) was selected from the GALAH DR3 catalogue \citep{bud21} with Galactic longitude $l \geq$ 330 degrees and Galactic latitude $12 < |b| <$ 30 degrees, and cross-matched with Gaia EDR3 and Skymapper to retrieve their photometric and astrometric information. The data set for our simulation includes 10500 stars with available spectra from the GALAH database\footnote{\url{https://datacentral.org.au/services/download}}. The spectra of these 10500 stars were analysed with the \textsc{EPIC} method, which was developed and optimised for determination of stellar parameters of Sun-like stars (SDST I).

Based on the selected subset of bright GALAH stars, we aim to generate many realisations of simulated data sets with fainter magnitude range (15.4--17.4\,mag) and similar reddening in our target field. More specifically, the distributions of photometric/astrometric properties of each realisation's stars, i.e. G magnitude, reddening, parallax and distance, are simulated to mimic stars with the fainter magnitude range of our main targets, while realistic uncertainties for these quantities are drawn from a subset of 100--200 stars in that magnitude range in the Gaia EDR3 catalogue. The details are described in steps 1--4 below. \\
1. For the simulated data set, the magnitude of a given star is randomly drawn from the realistic magnitude distribution between 15.4 -- 17.4\,mag, based on the Gaia measurements in our target field. \\
2. The reddening value E(B $-$ V) of each simulated star is randomly drawn from the realistic reddening distribution in our target field, based on \citet{sch98} with revised coefficients \citep{sf11}. \\
3. The Gaia colours and absolute magnitude of a given star remain the same in principle. Therefore a simulated star with fainter magnitude would correspond to a smaller parallax (or a larger distance) as:
\begin{equation}
{\rm parallax}\_{\rm sim} = 10^{0.2\cdot(M_G - A_G + A_{G,\rm sim} + 10 - G_{\rm sim})}
\end{equation}
\begin{equation}
{\rm distance}\_{\rm sim} = \frac{1000}{{\rm parallax}\_{\rm sim}}
\end{equation}
where the unit of distance is in $pc$ and the unit of parallax is in $mas$. \\
4. With the simulated G magnitude and corresponding parallax and distance, scatter can be added to the stellar photometric and astrometric properties based on realistic uncertainties. For each given simulated star, a subset of stars ($>$\,100) with very similar magnitude, parallax, and reddening from the Gaia catalogue is chosen in our target field. We then randomly pick a star from this subset to fetch its representative and realistic property uncertainties, also from the Gaia catalogue. The added scatter to each simulated stellar property (e.g., G, BP, RP, parallax etc.) is then randomly drawn from a Gaussian distribution based on the corresponding uncertainties of the picked star in the chosen subset. \\

\begin{figure}
\centering
\includegraphics[width=\columnwidth]{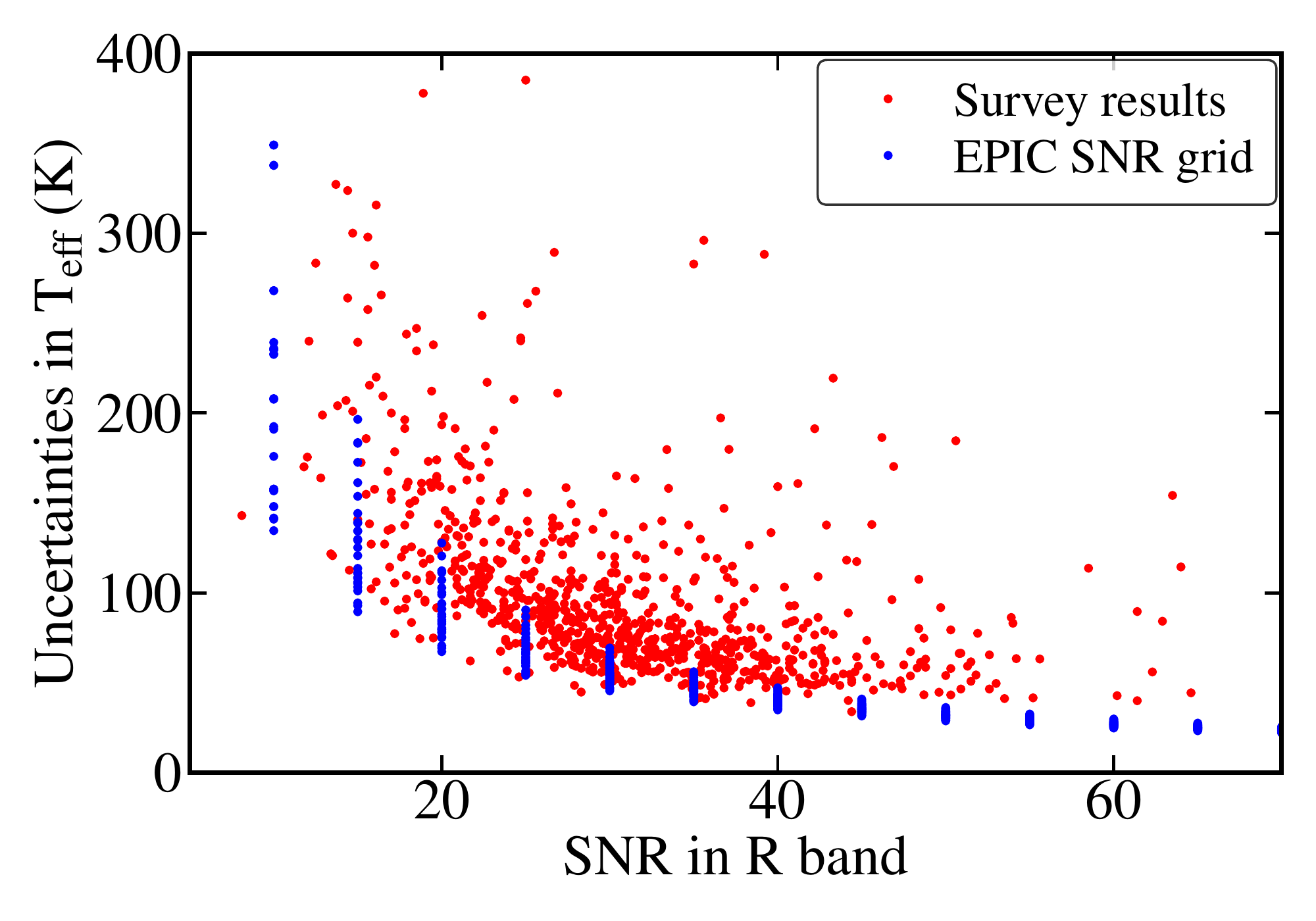}
\caption{\textsc{EPIC} uncertainties in \teff\ against SNR in the HERMES red band. The red and blue points represent the realistic uncertainties derived from the AAT/HERMES data for our survey and the expected uncertainties from the SNR grid based on a set of simulated solar spectra.}
\label{fig:err}
\end{figure}

It is also useful to establish spectroscopic parameters and their realistic uncertainties of this simulated data set with fainter magnitudes, in order to test our selection criteria with expected AAT exposure time and the corresponding HERMES spectral SNR. In principle, the spectroscopic parameters (derived by \textsc{EPIC}) of a realisation's stars are assumed to be unchanged, while their realistic uncertainties are drawn from an empirical grid based on the estimated spectral SNR in the four HERMES bands as functions of apparent magnitude and exposure time. The details are described in steps 5--6 below. \\
5. For each simulated star, the expected SNR in the four HERMES bands for the proposed exposure time (23 effective exposures with an average of 4056\,sec per exposure) can be estimated using the exposure calculation formula for AAT/HERMES \citep{she15}. To use this formula, we need to transform the simulated Gaia magnitude to Johnson passbands (B, V, R, and I) following the Gaia EDR3 documentation\footnote{\url{https://gea.esac.esa.int/archive/documentation/GEDR3/Data\_processing/chap\_cu5pho/cu5pho\_sec\_photSystem/cu5pho\_ssec\_photRelations.html}}. The sky brightness estimate for Siding Spring Observatory\footnote{\url{https://www.mso.anu.edu.au/pfrancis/reference/reference/node4.html}}, the total read and dark noise estimates from effectively 4.5 spatial pixels (which contribute each spectral pixel, see \citealp{kos17}) have been taken into account for our calculation. For a single 4056\,sec exposure, we expect e.g, SNR in the red band to be $\approx$\,6.2 per pixel for our faintest Sun-like candidates, and to be $\approx$\,29.6 per pixel for their combined spectra. The red dotted line in Fig.~\ref{fig:snr} shows the expected SNR in the HERMES red band for our simulated data set. \\
6. The spectroscopic parameters of the selected subset of GALAH stars were derived by \textsc{EPIC} for our simulated data set, while the parameter uncertainties were re-determined based on the expected spectral SNR in the four HERMES bands from the previous step. For this purpose, we made use of \textsc{EPIC} to estimate uncertainties in stellar parameters for simulated solar spectra with a grid of SNR in the four HERMES bands, e.g., SNR in the red band from 5 to 70 with a step size of 5. Therefore the uncertainties in the stellar parameters of each simulated star can be derived using such a grid for interpolation. Fig.~\ref{fig:err} shows the expected \textsc{EPIC} uncertainties in \teff\ against SNR in the red band from the SNR grid, as well as realistic \textsc{EPIC} uncertainties measured from our reduced AAT/HERMES data (see details in SDST III). There is a range of \teff\ uncertainties for a given SNR in the HERMES red band, which is due to the range of SNRs possible in the other three HERMES bands. We note the expected uncertainties are generally smaller than that of the measured uncertainties, which we will discuss in Section 4.3.3. \\

To test the effectiveness of our selection of faint solar analogues, the same selection criteria (Quality control, Gaia photometric and astrometic criteria, and Skymapper criteria) were applied to the simulated data set (generated with above steps 1--6) for each realisation. The expected success rate for this set of data can then be derived by obtaining the fraction of selected stars belonging to our definition of solar analogues out of number of stars fulfilling our selection criteria. 

Overall we simulated 10,000 realisations in order to reliably estimate the success rate in selecting faint solar analogues in our target field, which allowed us to confidently plan for our HERMES observations as stated in Section 3. The estimated success rate as a function of Gaia G magnitude is shown in Fig.~\ref{fig:success} (black dashed curve). This is considered as the preliminary expectation, where the success rate gradually decrease from $\approx$ 60\% to 50\% with increasing G magnitudes. Fig.~\ref{fig:success} also shows the results based on the stellar population model (green dashed line) and the realistic measurements (orange and magenta lines), demonstrating that we managed to identify more than 150 solar analogues as expected. Further explanation and discussion are presented in below subsections.

\subsubsection{Success rate based on stellar population model}

As stated, it is important to examine whether the number of potential solar analogues is reasonable in our selected target field to add confidence to the preliminary expectation derived above. Therefore theoretical stellar population models were also taken into account to better estimate the selection success rate of solar analogues. with similar methodology as presented in Section 4.3.1. 

In this work, the Besan\c{c}on stellar population model \citep{cze14,rob12} was adopted to provide a synthetic catalogue of stars in the same 2-degree target field. The colours (B-V, B-I, V-R and V-I) from the Besan\c{c}on model were transformed to Gaia G, BP-RP, BP-G0 and G0-RP using the equations from the Gaia EDR3 documentation. We ensured that the reddening distribution from the model is very similar to that from our target field with an average E(B $-$ V) of 0.075 and a standard deviation of 0.010. 

About 13500 stars with Gaia G between 15.4 -- 17.4\,mag were selected for this test. The success rate was estimated based on the Besan\c{c}on model with a similar approach as presented in Step\,4 of Section 4.3.1, to obtain realistic uncertainties in model stars' parameters (e.g., G, BP, RP, distance etc.) and to add scatter to these parameters based on their realistic uncertainties. The same selection criteria were then applied to estimate the success rate as described in Section 4.3.1. The Skymapper criteria was excluded as there is no robust transformation of colours using the very narrow Skymapper $v$ band to any Johnson colours used in the Besan\c{c}on model. Finally we run the test for 10,000 times to reliably estimate the success rate and its distribution based on the Besan\c{c}on stellar population model. 

\begin{figure}
\centering
\includegraphics[width=\columnwidth]{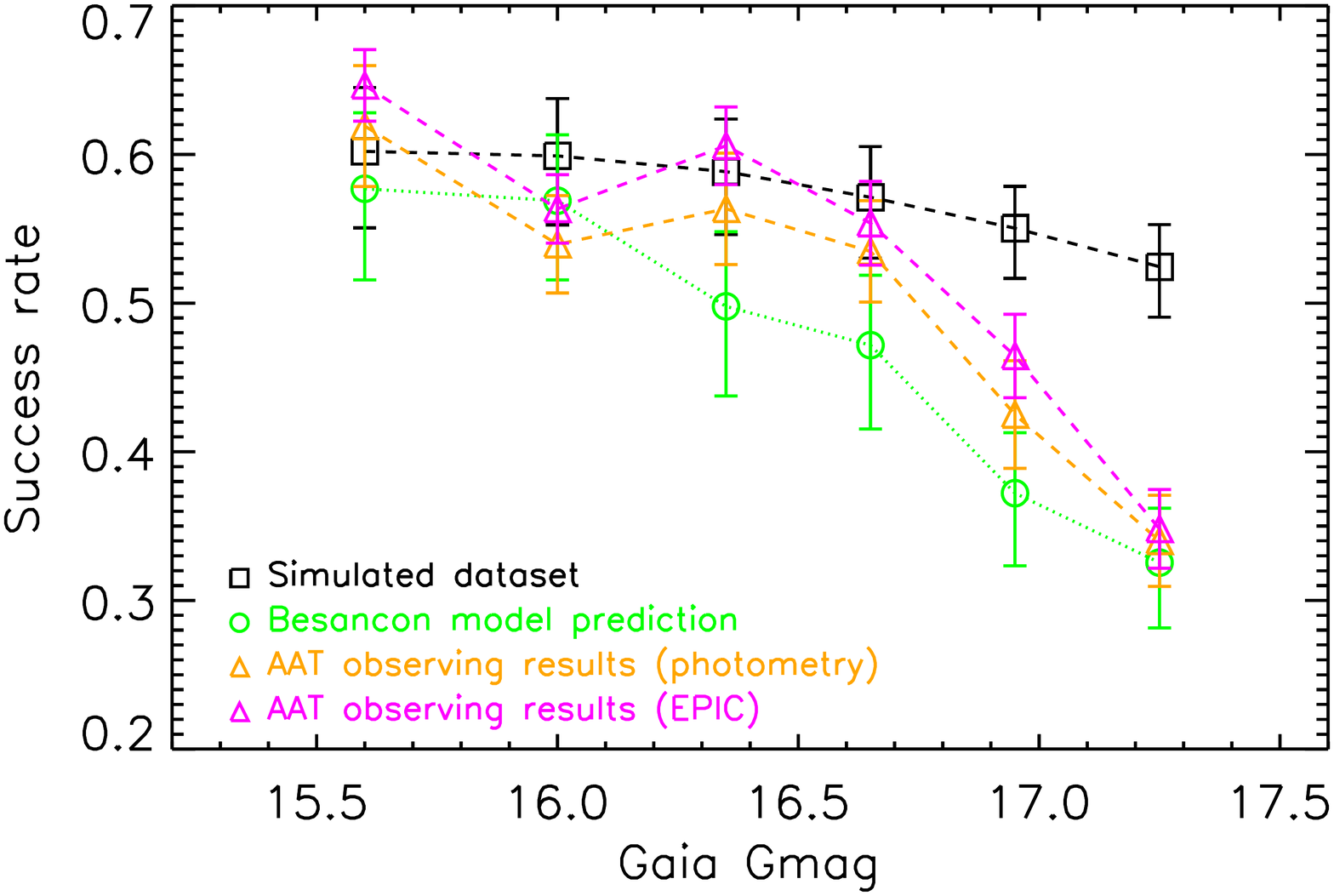}
\caption{Success rate of our selection as a function of Gaia G magnitudes. The black and green lines represent the success rate derived from our simulated data set and from the Besan\c{c}on model. The orange and magenta lines represent the success rate based on the stellar parameters derived using the photometric method (as described in Section 4.2) and the spectroscopic method (\textsc{EPIC}) as presented in SDST III.}
\label{fig:success}
\end{figure}

\begin{figure}
\centering
\includegraphics[width=\columnwidth]{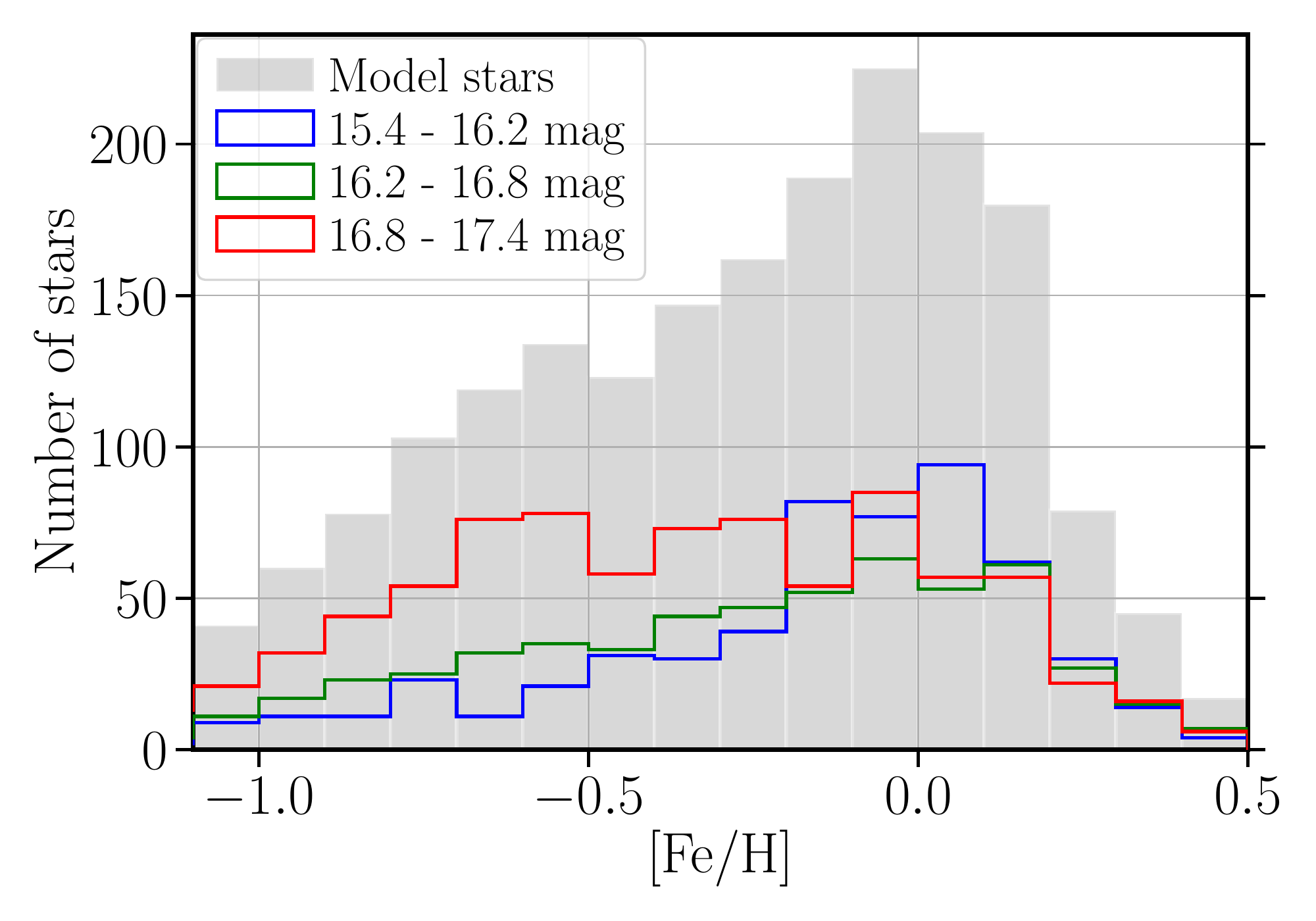}
\caption{Metallicity distribution of our target field based on the Besan\c{c}on model. The grey, blue, green, and red distribution represent all the model stars, the model stars falling into the magnitude bin of 15.4 -- 16.2\,mag, 16.2 -- 16.8\,mag, and 16.8 -- 17.4\,mag, respectively.}
\label{fig:hist3}
\end{figure}

As shown in Fig.~\ref{fig:success} (green dashed line), the success rate from the stellar population model decreases from $\approx$ 60\% to 30\% with increasing G magnitudes (up to 17.4\,mag), which is generally lower than that from the simulated data set. The difference becomes significant for fainter stars with G $>$ 16.8 (the faintest magnitude bin). The main reason is due to the selection of much more distant targets above the Galactic plane, which naturally leads to a shifted metallicity distribution \citep{hay15}. We show the metallicity distribution predicted by the Besan\c{c}on model in our target field for the three separated magnitude bins in Fig.~\ref{fig:hist3}. It is clear that the stars within the faintest magnitude bin have relatively lower metallicities with a skewed peak from 0.0\,dex towards $-$0.5\,dex. 

\subsubsection{Success rate based on observing results}

The success rate was derived based on our actual HERMES observing results and compared to those based on the preliminary expectation and the stellar population model. The results from SDST III were used to derive this spectroscopic success rate plotted in Fig.~\ref{fig:success} as the magenta points. In addition, the stellar parameters derived with photometric method (as described in Section 4.2) and their associated uncertainties were also used to calculate the expected success rate (orange dashed line in Fig.~\ref{fig:success}). 

Among the 547 main candidates, there are more than 150 stars within our definition as solar analogues (a further discussion is presented in SDST III). As shown in Fig.~\ref{fig:success}, we found the selection success rates of solar analogues based on the photometric and spectroscopic method agree well with each other, both decrease from $\approx$ 60\% to 30\% with increasing magnitude from 15.4 to 17.4. Although with some fluctuation, probably associated with binning (0.3\,mag per bin), the realistic success rates agree with that based on the theoretical model, especially for the faintest targets, indicating that we successfully selected and identified the expected number of solar analogues across the magnitude and distance range. 

We note the success rate estimates from observing results and the stellar population model are lower than those from preliminary expectations (mainly used for planning of observations), which is worth some further discussion to help improve future surveys of Sun-like stars. Several factors which may drive down the success rate from our preliminary expectation are discussed below, especially for the faintest stars. 
\begin{itemize}
\item The metallicity distribution of our candidates (especially the faintest ones) in the target field is skewed towards more metal-poor based on the Besan\c{c}on stellar population model, as stated in Section 4.3.2 and illustrated in Fig.~\ref{fig:hist3}. We expect this to be reflected in our real sample as well, skewing the metallicity distribution of our main candidates away from zero metallicity at larger distances (fainter magnitudes), thereby reducing the number of candidates which have metallicities within our solar analogue range ($\pm$0.3 dex). Given the results in Fig.~\ref{fig:hist3}, where we observe this effect directly for the Besancon model results, we attribute the majority of the drop in our real success rate at fainter magnitudes to this effect. 
\item The reddening E(B $-$ V) adopted in our survey was taken from the two-dimensional dust map from \citet{sch98} with revised coefficients \citep{sf11}. They may differ by a few 0.01 dex from the realistic reddening values if the dust distribution along the distance is taken into account. Unfortunately we note that the three-dimensional dust map (e.g., \citealp{gre19}) in our target sky region is not available. This will induce certain amount of systematics in terms of the \teff\ distribution of our selected candidates. For example, if the E(B $-$ V) values from two-dimensional dust map are systematically higher than the realistic values, this will lead us to select cooler stars (by 50 -- 100\,K) with systematically higher $(BP - RP)_0$, thus suppressing the success rate. This may explain the slightly cooler \teff\ distribution of our main targets. 
\item There is a range of spectral SNRs for candidates with very similar apparent magnitudes, with the distribution skewed to lower SNR compared to the predicted values from the exposure time calculation (see Fig.~\ref{fig:snr}). This is mostly due to the lower throughput of the corresponding fibres. As shown in Fig.~\ref{fig:err}, this also leads to underestimation of \textsc{EPIC} uncertainties in spectroscopic parameters from the SNR grid adopted in the simulated data set. This causes the preliminary success rate estimate to be overestimated, particularly at fainter magnitudes. 
\item In addition, the expected uncertainties in spectroscopic parameters based on a grid of SNR are lower than the measured uncertainties from our observations, as stated in Step\,6 in Section 4.3.1 and shown in Fig.~\ref{fig:err}. It is mainly because the \textsc{EPIC} method is based on differential analysis relative to the Sun, where the systematics increase when the stellar parameters deviate from the solar values. Therefore the expected uncertainties of the grid, purely based on the simulated solar spectra, should be systematically lower than that of our observing results with a range of stellar parameters around solar values, e.g., (\teff, \logg, [Fe/H]) within $\pm$(300\,K, 0.4\,dex, 0.3\,dex), or even with larger range. This, again, affects our preliminary estimate of success rate. 
\end{itemize}

We found that the selection success rate of solar analogues in our survey based on the Besan\c{c}on stellar population model agree well with that from the observing results, from spectroscopy (and also from the photometric method), even for the faintest candidates. Therefore, we can conclude that we are able to discover and confirm the expected number of solar analogues ($\gtrsim$\,150) in our target field. Many of these are much fainter (2.5\,mag deeper) and much more distant (2.5–-3\,kpc) than the vast majority of known solar twins/analogues\footnote{While the most distant solar twins confirmed, with high confidence, through spectroscopy are in M67 at distance 890\,pc \citep{kha13}, we note that \citet{ben17} have spectroscopically confirmed $\approx$9 microlensed solar analogues which likely reside in the Galactic Bulge.}. In this work, the stellar properties of our candidates were derived for initial verification, while the final spectroscopic identification of the most likely solar twins/analogues will be presented in SDST III. We plan to follow-up a sub-sample of the confirmed twins and analogues with VLT/ESPRESSO, which will allow us to measure precisely any variations in $\alpha$ closer to the Galactic Centre as a function of varying Dark Matter density.

\section{Summary}

The goal of the SDST is to select and identify distant solar twins and analogues as new probes to search for potential variations in the strength of electromagnetism, $\alpha$, across the Milky Way with varying Dark Matter density. \citet{ber22a,ber22b} and \citet{mur22a} have demonstrated that high-resolution spectra of stars similar to the Sun can be used to measure the fine-structure constant $\alpha$ at a precision of $\lesssim$30\,ppb per star, which is about two orders of magnitude better than current astronomical tests. In this paper we reported our selection, observations and data reduction of our initial Sun-like candidates with AAT/HERMES. In SDST III (Lehmann et al., in prep.) we will spectroscopically confirm the most Sun-like amongst them -- solar twins and analogues -- over a distance range up to 4\,kpc closer to the Galactic Centre for VLT/ESPRESSO follow-up. 

We made use of Gaia EDR3 and Skymapper catalogue to select our initial Sun-like targets in a 2-degree field closer to the Galactic Centre ($l \approx$ 30\,degrees) and above the Galactic plane ($b \approx$ 16.5\,degrees) for AAT/HERMES observations in the second semester of 2021. The candidates are required to have reliable Gaia and Skymapper photomeric measurements, low reddening with E(B $-$ V) $<$ 0.12, and relative uncertainties in Gaia parallaxes $<$ 50\%. After zero-point correction and dereddening, our initial Sun-like candidates were selected to have close-to-solar properties in the colour-magnitude diagram with additional criteria on Skymapper $(v-g)$ to remove stars with low metallicity ([Fe/H] $<$ $-$0.8). In total 1201 candidates were selected in our target field, with 877 of them observed during our AAT/HERMES run. 

Amongst the 877 Sun-like candidates, 547 are our main candidates with Gaia G between 15.4 to 17.4\,mag. We obtained 23 effective exposures with an average exposure time of 4056\,sec for the main targets as well as additional 5 effective exposures for the rest brighter targets (G $<$ 15.4\,mag). The spectra were reduced with customised \textit{2dfdr} pipeline, and adopted our own algorithm for sky subtraction and telluric correction (with the \textit{Molecfit} software). The reduced spectra of each object were combined together with proper weighting using \textit{2dfdr}. Our observing strategy successfully provided the required spectral SNR (e.g., $\geq$\,25 per pixel in the HERMES red band) across the magnitude range for the majority of the main candidates. We highlight that the SDST is about 2--2.5\,mag deeper than today's large spectroscopic surveys (e.g., Gaia-ESO, APOGEE, and GALAH), providing us completely new information of the most distant Sun-like stars (1--4\,kpc away) in the inner Galaxy. 

The stellar properties (e.g., \teff, \logg, mass) were derived using the photometric method as described in Section 4.2 for initial verification. We found $\gtrsim$ 92\% of sample stars are within our definition of solar analogues in terms of their \teff\ and \logg\ values. We then estimated, examined and compared the selection success rate of solar analogues, based on the simulated data set, the Besan\c{c}on stellar population model, and the observing results (obtained with both photometric method, and spectroscopic method \textsc{EPIC} as presented in SDST III). We are able to achieve a realistic success rate from $\approx$ 60\% to 30\% for stars with G magnitude from 15.4 to 17.4. We emphasize that the estimation of success rate based on the Besan\c{c}on model agrees well with that from the observing results (as shown in Fig.~\ref{fig:success}), which provides us confidence to conclude that we successfully selected and identified $\gtrsim$\,150 solar analogues, as expected, across the magnitude and distance range. 

As discussed in Section 4.3.3, the realistic success rates are lower than that from our preliminary expectation, which is mainly due to the skewed metallicity distribution for the most distant targets towards $-$0.5 dex, as revealed by our analysis of the Besan\c{c}on model results (Fig.~\ref{fig:hist3}). Other factors affecting the estimate of success rate include, e.g., systematic overestimate of the reddening, E(B $-$ V), due to the use of a two-dimensional dust map; underestimated uncertainties in spectroscopic parameters at a given magnitude due to unexpected lower spectral SNR for a number of candidates (because of lower fibre throughput), as well as the underestimation of \textsc{EPIC} uncertainties drawn from a grid of SNR, based purely on a set of simulated solar spectrum. Future surveys of Sun-like stars may benefit by taking into account the above factors, most importantly, the metallicity distribution in the different regions of our galaxy.

In this paper we provide the final reduced 1-dimensional AAT/HERMES spectra of the SDST, along with the stellar photometric/astrometric properties of our sample stars. The final spectroscopic stellar parameters, derived using EPIC, will be presented in SDST III. The selection success rates of solar analogues in our survey agree well between theoretical prediction and observations, and demonstrate that we are able to select, analyse, and identify solar analogues much fainter (2--2.5\,mag deeper) and much more distant (2.5--3\,kpc further) than the local sample of solar twins/analogues. SDST III will also present the most likely solar twins and analogues that can be followed-up using the ultra-stable ESPRESSO spectrograph on the VLT. This will provide a new measure of any changes in $\alpha$ as a function of distance -- and, therefore, dark matter density -- across our galaxy. In addition, the AAT/HERMES and VLT/ESPRESSO spectra of these new, distant solar twins and analogues will enable us to measure (differential) elemental abundances of more than 20 elements at a $\approx$\,0.05--0.1\,dex precision level. This could provide new insights into Galactic chemical evolution, nucleosynthetic processes, star formation history, and serve as a new reference sample for, e.g., Galactic Archaeology in the inner region of our Milky Way.

\section*{Acknowledgements}
FL, MTM and CL acknowledge the support of the Australian Research Council through Future Fellowship grant FT180100194. JK acknowledges financial support of the Slovenian Research Agency (research core funding No. P1-0188). SLM acknowledges the support of the Australian Research Council through Discovery Project grant DP180101791, and the support of the UNSW Scientia Fellowship Program.\\ 
This work has made use of data from the European Space Agency (ESA) mission {\it Gaia} (\url{https://www.cosmos.esa.int/gaia}), processed by the {\it Gaia} Data Processing and Analysis Consortium (DPAC, \url{https://www.cosmos.esa.int/web/gaia/dpac/consortium}). Funding for the DPAC has been provided by national institutions, in particular the institutions participating in the {\it Gaia} Multilateral Agreement.\\
The national facility capability for SkyMapper has been funded through ARC LIEF grant LE130100104 from the Australian Research Council, awarded to the University of Sydney, the Australian National University, Swinburne University of Technology, the University of Queensland, the University of Western Australia, the University of Melbourne, Curtin University of Technology, Monash University and the Australian Astronomical Observatory. SkyMapper is owned and operated by The Australian National University's Research School of Astronomy and Astrophysics.\\
This work is based on data acquired at the Anglo-Australian Telescope, under program A/2021A/005. We acknowledge the traditional custodians of the land on which the AAT stands, the Gamilaraay people, and pay our respects to elders past and present.\\
We acknowledge the use of NASA's SkyView facility (http://skyview.gsfc.nasa.gov) located at NASA Goddard Space Flight Center. We acknowledge our usage of \textit{TOPCAT} software \citep{tay05} for this work. We acknowledge the Python Software Foundation and our usage of the Python software packages \textit{Astropy, Matplotlib, Numpy, Pandas}, and \textit{Scipy}. 

\section*{Data Availability}
The raw and reduced spectral data underlying this article are available at  \url{https://datacentral.org.au} under AAT program A/2021A/005. The additional GALAH spectral data underlying this article are also available at \url{https://datacentral.org.au/services/download}. The rest of data underlying this article are available in the article and in its online supplementary material.

\bibliographystyle{mnras}

\begin{thebibliography}{99}
\bibitem[\protect\citeauthoryear{AAO Software Team}{2015}]{aao15} AAO Software Team, 2015, Astrophysics Source Code Library, record ascl:1505.015
\bibitem[\protect\citeauthoryear{Aoyama et al.}{2015}]{aoy15} Aoyama T., Hayakawa M., Kinoshita T., Nio M., 2015, PhRvD, 91, 033006
\bibitem[\protect\citeauthoryear{Adibekyan et al.}{2021}]{adi21} Adibekyan V., et al. 2021, Science, 374, 330
\bibitem[\protect\citeauthoryear{Asplund et al.}{2009}]{asp09} Asplund M., Grevesse N., Sauval A.~J., Scott P., 2009, \araa, 47, 481
\bibitem[\protect\citeauthoryear{Bailer-Jones et al.}{2021}]{bai21} Bailer-Jones C.~A.~L., Rybizki J., Fouesneau M., Demleitner M., Andrae R., 2021, \aj, 161, 147
\bibitem[\protect\citeauthoryear{Baumann et al.}{2010}]{bau10} Baumann P., Ram\'irez I., Mel\'endez J., Asplund M., Lind K., 2010, \aap, 519, A87
\bibitem[\protect\citeauthoryear{Bedell et al.}{2018}]{bed18} Bedell M., et al., 2018, \apj, 865, 68
\bibitem[\protect\citeauthoryear{Bensby et al.}{2017}]{ben17} Bensby T., et al., 2017, \aap, 605, A89
\bibitem[\protect\citeauthoryear{Bensby et al.}{2014}]{ben14} Bensby T., Feltzing S., Oey M.~S., 2014, \aap, 562, A71
\bibitem[\protect\citeauthoryear{Berke et al.}{2022a}]{ber22a} Berke D.~A, Murphy M.~T., Flynn C., Liu F., 2022a, submitted to MNRAS
\bibitem[\protect\citeauthoryear{Berke et al.}{2022b}]{ber22b} Berke D.~A, Murphy M.~T., Flynn C., Liu F., 2022b, submitted to MNRAS
\bibitem[\protect\citeauthoryear{Botelho et al.}{2020}]{bot20} Botelho R.~B., Milone A.~C., Mel\'endez J., Alves-Brito A., Spina L., Bean J.~L., 2020, \mnras, 499, 2196
\bibitem[\protect\citeauthoryear{Buder et al.}{2021}]{bud21} Buder S., et al., 2021, \mnras, 506, 150
\bibitem[\protect\citeauthoryear{Carlos et al.}{2019}]{car19} Carlos M., et al., 2019, \mnras, 485, 4052
\bibitem[\protect\citeauthoryear{Carlos, Nissen \& Mel\'endez}{2016}]{car16} Carlos M., Nissen P. E., Mel\'endez J., 2016, \aap, 587, A100
\bibitem[\protect\citeauthoryear{Casagrande et al.}{2021}]{cas21} Casagrande L., et al., 2021, \mnras, 507, 2684
\bibitem[\protect\citeauthoryear{Casagrande et al.}{2019}]{cas19} Casagrande L., Wolf C., Mackey A.~D., Nordlander T., Yong D., Bessell M., 2019, \mnras, 482, 2770
\bibitem[\protect\citeauthoryear{Casagrande \& VandenBerg}{2018}]{cv18} Casagrande L., VandenBerg D.~A., 2018, \mnras, 479, L102
\bibitem[\protect\citeauthoryear{Casagrande et al.}{2010}]{cas10} Casagrande L., Ram\'irez I., Mel\'endez J., Bessell M., Asplund M., 2010, \aap, 512, A54
\bibitem[\protect\citeauthoryear{Casali et al.}{2020}]{cas20} Casali G., et al., 2020, \aap, 639, A127
\bibitem[\protect\citeauthoryear{Cayrel de Strobel}{1996}]{cay96} Cayrel de Strobel G., 1996, A\&ARv, 7, 243
\bibitem[\protect\citeauthoryear{Cayrel de Strobel et al.}{1981}]{cay81} Cayrel de Strobel G., Knowles N., Hernandez G., Bentolila C., 1981, \aap, 94, 1
\bibitem[\protect\citeauthoryear{Czekaj et al.}{2014}]{cze14} Czekaj M.~A., Robin A.~C., Figueras F., Luri X., Haywood M., 2014, \aap, 564, A102
\bibitem[\protect\citeauthoryear{Datson, Flynn \& Portinari}{2015}]{dat15} Datson J., Flynn C., Portinari L., 2015, \aap, 574, A124
\bibitem[\protect\citeauthoryear{Datson, Flynn \& Portinari}{2014}]{dat14} Datson J., Flynn C., Portinari L., 2014, \mnras, 439, 1028
\bibitem[\protect\citeauthoryear{Datson, Flynn \& Portinari}{2012}]{dat12} Datson J., Flynn C., Portinari L., 2012, \mnras, 426, 484
\bibitem[\protect\citeauthoryear{Davoudiasl \& Giardino}{2019}]{dav19} Davoudiasl H., Giardino P.~P., 2019, Phys.Lett.B, 788, 270
\bibitem[\protect\citeauthoryear{Demarque et al.}{2004}]{dem04} Demarque P., Woo J.-H., Kim Y.-C., Yi S.-K., 2004, ApJS, 155, 667
\bibitem[\protect\citeauthoryear{De Silva et al.}{2015}]{des15} De Silva G.~M., et al., 2015, \mnras, 449, 2604
\bibitem[\protect\citeauthoryear{Eichhorn, Held \& Wetterich}{2018}]{eic18} Eichhorn A., Held A., Wetterich C., 2018, Phys.Lett.B, 782, 198
\bibitem[\protect\citeauthoryear{Friel et al.}{1993}]{fri93} Friel E., Cayrel de Strobel G., Chmielewski Y., Spite M., Lebre A., Bentolila C., 1993, \aap, 274, 825
\bibitem[\protect\citeauthoryear{Gaia Collaboration et al.}{2021}]{gaia21} Gaia Collaboration, et al., 2021, \aap, 649, A1
\bibitem[\protect\citeauthoryear{Gaia Collaboration et al.}{2016}]{gaia16} Gaia Collaboration, et al., 2016, \aap, 595, A1
\bibitem[\protect\citeauthoryear{Green et al.}{2019}]{gre19} Green G.~M., Schlafly E., Zucker C., Speagle J.~S., Finkbeiner D., 2019, \apj, 887, 93
\bibitem[\protect\citeauthoryear{Gustafsson}{2018}]{gus18} Gustafsson B., 2018, \aap, 616, A91
\bibitem[\protect\citeauthoryear{Hanneke, Fogwell \& Gabrielse}{2008}]{han08} Hanneke D., Fogwell S., Gabrielse G., 2008, PhRvL, 100, 120801
\bibitem[\protect\citeauthoryear{Hardorp}{1978}]{har78} Hardorp J., 1978, \aap, 63, 383
\bibitem[\protect\citeauthoryear{Hayden et al.}{2015}]{hay15} Hayden M.~R., et al., 2015, \apj, 808, 132 
\bibitem[\protect\citeauthoryear{Hees et al.}{2020}]{hee20} Hees A., et al., 2020, PhRvL, 124, 081101
\bibitem[\protect\citeauthoryear{Holmberg, Flynn \& Portinari}{2006}]{hol06} Holmberg J., Flynn C., Portinari L., 2006, \mnras, 367, 449
\bibitem[\protect\citeauthoryear{Kharchenko et al.}{2013}]{kha13} Kharchenko N.~V., Piskunov A.~E., Schilbach E., R\"oser S., Scholz R.~D., 2013, \aap, 558, A53
\bibitem[\protect\citeauthoryear{Kausch et al.}{2015}]{kau15} Kausch W., et al., 2015, \aap, 576, A78
\bibitem[\protect\citeauthoryear{Keller et al.}{2007}]{kel07} Keller S.~C., et al., 2007, PASA, 24, 1
\bibitem[\protect\citeauthoryear{King, Boesgaard \& Schuler}{2005}]{kin05} King J.~R., Boesgaard A. M., Schuler S.~C., 2005, \aj, 130, 2318
\bibitem[\protect\citeauthoryear{Kos et al.}{2017}]{kos17} Kos J., et al., 2017, \mnras, 464, 1259
\bibitem[\protect\citeauthoryear{Lange et al.}{2021}]{lan21} Lange R., et al., 2021, PhRvL, 126, 011102
\bibitem[\protect\citeauthoryear{Lehmann et al.}{2022}]{leh22a} Lehmann C., Murphy M.~T., Liu F., Flynn C., Berke D.~A, 2022, \mnras, 512, 11
\bibitem[\protect\citeauthoryear{Lewis et al.}{2002}]{lew02} Lewis I.~J., et al., 2002, \mnras, 333, 279
\bibitem[\protect\citeauthoryear{Lindegren et al.}{2021}]{lin21} Lindegren L., et al., 2021, \aap, 649, A4
\bibitem[\protect\citeauthoryear{Liu et al.}{2020}]{liu20} Liu F., Yong D., Asplund M., Wang H.~S., Spina L., Acu\~{n}a L., Mel\'endez J., Ram\'irez I., 2020, \mnras, 495, 3961
\bibitem[\protect\citeauthoryear{Liu et al.}{2016}]{liu16} Liu F., Asplund M., Yong D., Mel\'endez J., Ram\'irez I., Karakas A.~I., Carlos M., Marino A.~F., 2016, \mnras, 463, 696
\bibitem[\protect\citeauthoryear{Lorenzo-Oliveira et al.}{2018}]{oli18} Lorenzo-Oliveira D., et al., 2018, \aap, 619, A73
\bibitem[\protect\citeauthoryear{Mel\'endez et al.}{2014a}]{mel14a} Mel\'endez J., Schirbel L., Monroe T.~R., Yong D., Ram\'irez I., Asplund M., 2014a, \aap, 567, L3
\bibitem[\protect\citeauthoryear{Mel\'endez et al.}{2014b}]{mel14b} Mel\'endez J., et al., 2014b, \apj, 791, 14
\bibitem[\protect\citeauthoryear{Mel\'endez et al.}{2009}]{mel09} Mel\'endez J., Asplund M., Gustafsson B., Yong D., 2009, \apjl, 704, L66
\bibitem[\protect\citeauthoryear{Mel\'endez \& Ram\'irez}{2007}]{mr07} Mel\'endez J., Ram\'irez I., 2007, \apj, 669, L89
\bibitem[\protect\citeauthoryear{Mel\'endez, Dodds-Eden \& Robles}{2006}]{mel06} Mel\'endez J., Dodds-Eden K., Robles J.~A., 2006, \apj, 641, L133
\bibitem[\protect\citeauthoryear{Murphy et al.}{2022a}]{mur22a} Murphy M.~T., Berke D.~A., Liu F., Flynn C., Lehmann C., Dzuba V.~A., Flambaum V.~V., 2022a, submitted to Science
\bibitem[\protect\citeauthoryear{Murphy et al.}{2022b}]{mur22b} Murphy M.~T., et al., 2022b, \aap, 658, A123 
\bibitem[\protect\citeauthoryear{Murphy \& Cooksey}{2017}]{mur17} Murphy M.~T, Cooksey K.~L., 2017, \mnras, 471, 4930
\bibitem[\protect\citeauthoryear{Nissen et al.}{2020}]{nis20} Nissen P.~E., Christensen-Dalsgaard J., Mosumgaard J.~R., Silva Aguirre V., Spitoni E., Verma K., 2020, \aap, 640, A81
\bibitem[\protect\citeauthoryear{Nissen}{2015}]{nis15} Nissen P.~E., 2015, \aap, 579, A52
\bibitem[\protect\citeauthoryear{Olive \& Pospelov}{2002}]{oli02} Olive K.~A., Pospelov M., 2002, PhRvD, 65, 085044
\bibitem[\protect\citeauthoryear{\"Onehag et al.}{2011}]{one11} \"Onehag A., Korn A., Gustafsson B., Stempels E., Vandenberg D.~A., 2011, A\&A, 528, A85
\bibitem[\protect\citeauthoryear{Onken et al.}{2019}]{onk19} Onken C.~A., et al., 2019, PASA, 36, 
\bibitem[\protect\citeauthoryear{Pasquini et al.}{2008}]{pas08} Pasquini L., Biazzo K., Bonifacio P., Randich S., Bedin L.~R., 2008, \aap, 489, 677
\bibitem[\protect\citeauthoryear{Pepe et al.}{2021}]{pep21} Pepe F., et al., 2021, \aap, 645, A96
\bibitem[\protect\citeauthoryear{Porto de Mello et al.}{2014}]{por14} Porto de Mello G.~F., da Silva R., da Silva L., de Nader R.~V., 2014, \aap, 563, A52
\bibitem[\protect\citeauthoryear{Porto de Mello \& da Silva}{1997}]{pd97} Porto de Mello G.~F., da Silva L., 1997, \apj, 482, L89
\bibitem[\protect\citeauthoryear{Pr{\v{s}}a et al.}{2016}]{prs16} Pr{\v{s}}a A., et al., 2016, \aj, 152, 41
\bibitem[\protect\citeauthoryear{Ram\'irez et al.}{2014}]{ram14} Ram\'irez I., et al., 2014, \aap, 572, A48
\bibitem[\protect\citeauthoryear{Ram\'irez et al.}{2012}]{ram12} Ram\'irez I., et al., 2012, \apj, 752, 5
\bibitem[\protect\citeauthoryear{Ram\'irez et al.}{2010}]{ram10} Ram\'irez I., Asplund M., Baumann P., Mel\'endez J., Bensby T., 2010, \aap, 521, A33
\bibitem[\protect\citeauthoryear{Ram\'irez, Mel\'endez \& Asplund}{2009}]{ram09} Ram\'irez I., Mel\'endez J., Asplund M., 2009, \aap, 508, L17
\bibitem[\protect\citeauthoryear{Riello et al.}{2021}]{rie21} Riello M., et al., 2021, \aap, 649, A3
\bibitem[\protect\citeauthoryear{Robin et al.}{2012}]{rob12} Robin A.~C., Marshall D.~J., Schultheis M., Reyl\'e C., 2012, \aap, 538, A106
\bibitem[\protect\citeauthoryear{Rosenband et al.}{2008}]{ros08} Rosenband T., et al., 2008, Science, 319, 1808
\bibitem[\protect\citeauthoryear{Schlafly \& Finkbeiner}{2011}]{sf11} Schlafly E.~F., Finkbeiner D.~P., 2011, \apj, 737, 103
\bibitem[\protect\citeauthoryear{Schlegel, Finkbeiner \& Davis}{1998}]{sch98} Schlegel D.~J., Finkbeiner D.~P., Davis M., 1998, \apj, 500, 525
\bibitem[\protect\citeauthoryear{Sharp \& Birchall}{2010}]{sb10} Sharp R., Birchall M.~N., 2010, PASA, 27, 91
\bibitem[\protect\citeauthoryear{Sheinis et al.}{2015}]{she15} Sheinis A., et al., 2015, J. Astron. Telesc. Instrum. Syst., 1, 035002
\bibitem[\protect\citeauthoryear{Smette et al.}{2015}]{sme15} Smette A., et al., 2015, \aap, 576, A77
\bibitem[\protect\citeauthoryear{Spina et al.}{2018}]{spi18} Spina L., et al., 2018, \mnras, 474, 2580
\bibitem[\protect\citeauthoryear{Spina et al.}{2016}]{spi16} Spina L., Mel\'endez J., Karakas A.~I., Ram\'irez I., Monroe T.~R., Asplund M., Yong D., 2016, \aap, 593, A125
\bibitem[\protect\citeauthoryear{Stadnik \& Flambaum}{2015}]{sta15} Stadnik Y.~V., Flambaum V.~V., 2015, PhRvL, 115, 201301
\bibitem[\protect\citeauthoryear{Taylor}{2005}]{tay05} Astronomical Data Analysis Software and Systems XIV, eds. P Shopbell et al., 2005, ASP Conf. Ser. 347, p.~29
\bibitem[\protect\citeauthoryear{Wang et al.}{2019}]{wang19} Wang H.~S., Liu F., Ireland T.~R., Brasser R., Yong D., Lineweaver C.~H., 2019, \mnras, 482, 2222
\bibitem[\protect\citeauthoryear{Yana Galarza et al.}{2021}]{gal21} Yana Galarza et al., 2021, \mnras, 504, 1873
\bibitem[\protect\citeauthoryear{Yana Galarza et al.}{2016}]{gal16} Yana Galarza J., Mel\'endez J., Ram\'irez I., Yong D., Karakas A.~I., Asplund M., Liu F., 2016, \aap, 589, A17
\end{thebibliography}




\section*{Supplementary material}
The following supplementary material is available for this
article online: full version of Table~2.

\bsp	
\label{lastpage}
\end{CJK*}
\end{document}